\documentclass[sn-mathphys-num]{sn-jnl}

\usepackage[markup=underlined, defaultcolor=red]{changes}
\usepackage{graphicx}%
\usepackage{multirow}%
\usepackage{amsmath,amssymb,amsfonts}%
\usepackage{amsthm}%
\usepackage{mathrsfs}%
\usepackage[title]{appendix}%
\usepackage{xcolor}%
\usepackage{textcomp}%
\usepackage{manyfoot}%
\usepackage{booktabs}%
\usepackage{algorithm}%
\usepackage{algorithmicx}%
\usepackage{algpseudocode}%
\usepackage{listings}%
\usepackage{graphicx}
\usepackage{color}
\usepackage[english]{babel}
\usepackage{rotating}
\usepackage{braket}
\usepackage{multirow}
\usepackage{makecell}
\usepackage{longtable}
\usepackage{amsmath}
\usepackage{amssymb}
\usepackage{hyperref}
\usepackage{bm}
\usepackage{soul}
\usepackage{accents}
\usepackage{float}
\usepackage{appendix}
\usepackage{relsize}
\usepackage{booktabs}  
\usepackage{multirow}  
\usepackage{siunitx}   
\usepackage{pdfpages}

\newcommand{\abs}[1]{\left \lvert #1 \right \rvert}

\newcommand{\aop}{\hat a}

\newcommand{\bop}{\hat b}

\newcommand{\figref}[1]{\mbox{Fig.~\ref{#1}}}

\newcommand{\secref}[1]{\mbox{Sec.~\ref{#1}}}

\renewcommand{\eqref}[1]{\mbox{Eq.~(\ref{#1})}}

\newcommand{\expec}[1]{\left\langle{#1}\right\rangle}
\newcommand{\be}{\begin{equation}}
\newcommand{\ee}{\end{equation}}
\newcommand{\bea}{\begin{eqnarray}}
\newcommand{\eea}{\end{eqnarray}}

\definecolor{darkgreen}{rgb}{0.1, 0.3, 0.0}

\newcommand{\figpanel}[2]{Fig.~\hyperref[#1]{\ref*{#1}(#2)}}
\newcommand{\figpanels}[3]{Fig.~\hyperref[#1]{\ref*{#1}(#2)-(#3)}}
\newcommand{\figpanelNoPrefix}[2]{\hyperref[#1]{\ref*{#1}(#2)}}

\theoremstyle{thmstyleone}%
\theoremstyle{thmstyletwo}%

\theoremstyle{thmstylethree}%

\graphicspath{{./img/}}

\begin{document}

\title[Article Title]{Observation of Perfect Absorption in Hyperfine Levels of Molecular Spins with Hermitian Subspaces}

\author*[1,2]{\fnm{Claudio} \sur{Bonizzoni}}\email{claudio.bonizzoni@unimore.it}
\author[3]{\fnm{Daniele} \sur{Lamberto}}
\author[3]{\fnm{Samuel} \sur{Napoli}}
\author[4]{\fnm{Simon} \sur{G\"unzler}}
\author[4]{\fnm{Dennis} \sur{Rieger}}
\author[5]{\fnm{Fabio} \sur{Santanni}}
\author[2]{\fnm{Alberto} \sur{Ghirri}}
\author[4]{\fnm{Wolfgang} \sur{Wernsdorfer}}
\author[3]{\fnm{Salvatore} \sur{Savasta}}
\author[1,2]{\fnm{Marco} \sur{Affronte}}

\affil*[1]{\orgdiv{Dipartimento di Scienze Fisiche, Informatiche e Matematiche}, \orgname{Universit\`a di Modena e Reggio Emilia}, \orgaddress{\street{via G. Campi 213/a}, \city{Modena}, \postcode{41125}, \country{Italy}}}

\affil[2]{\orgdiv{Istituto Nanoscienze}, \orgname{CNR}, \orgaddress{\street{via G. Campi 213/a}, \city{Modena}, \postcode{41125}, \country{Italy}}}

\affil[3]{\orgdiv{Dipartimento di Scienze Matematiche e Informatiche, Scienze Fisiche e Scienze della Terra}, \orgname{Università degli Studi di Messina}, \orgaddress{\street{Via Salita Sperone}, \city{c.da Papardo, Messina}, \postcode{98166}, \country{Italy}}}

\affil[4]{\orgdiv{Physikalisches Institut}, \orgname{Karlsruhe Institute of Technology}, \orgaddress{\street{Wolfgang-Gaede-Str. 1}, \city{Karlsruhe}, \postcode{D-76131}, \country{Germany}}}

\affil[5]{\orgdiv Dipartimento di Chimica Ugo Schiff, \orgname{Università degli Studi di Firenze}, \orgaddress{\street{via della Lastruccia 3}, \city{Sesto Fiorentino (FI)}, \postcode{50019}, \country{Italy}}}

\date{\today}

\abstract{
We investigate Perfect Absorption (PA) of radiation, in which incoming energy is entirely dissipated, in a system consisting of molecular spin centers coherently coupled to a planar microwave resonator operated at milliKelvin temperature and in the single photon regime. This platform allows us to fine tune the spin-photon coupling and to control the effective dissipation of the two subsystems towards the environment, thus giving us the opportunity to span over a wide space of parameters. Our system can be effectively described by a non-Hermitian Hamiltonian exhibiting distinct Hermitian subspaces.
We experimentally show that these subspaces, linked to the presence of PA, can be engineered through the resonator-spin detuning, which controls the composition of the polaritons in terms of photon and spin content. In such a way, the required balance between the feeding and the loss rates is effectively recovered even in the absence of PT-symmetry.
We show that Hermitian subspaces influence the overall aspect of coherent spectra of cavity QED systems and enlarge the possibility to explore non-Hermitian effects in open quantum systems.
We finally discuss how our results can be potentially exploited for applications, in particular as single-photon switches and modulators.}


\maketitle

\section{Introduction}\label{sec_intro}

Passive open light-matter quantum systems offer the opportunity to investigate non-Hermitian physics \cite{sunPRL2014,wennerPRL2014,ZhenNAT2015,zhangNATCOMM2017}. Non-Hermiticity provides an insightful theoretical framework to understand unconventional phenomena such as wave-scattering anomalies, \textit{e.g.}, Coherent Perfect Absorption (CPA) \cite{WanScience2011, Zanotto2014Nature, wangSCIENCE2021}, Perfect Absorption (PA) (often also referred to as Reflectionless Scattering Modes, RSM, in single-port configuration) \cite{SweeneyPhysRevA.102.063511,FerisePRL_2022,JiangNature2024}, and electromagnetically-induced transparency \cite{Safavi-NaeiniNAT2011,qianNATCOMM2023}, which have attracted a great deal of interest in the last decades for their potential photonic applications \cite{feng2017non,wiersigPHOTRES2020}. These peculiar wave scattering effects are often associated with the presence of Exceptional Points (EP, \textit{i.e}, degeneracy of the complex resonance frequencies of the system) \cite{zhangPRXQUANTUM2021, wangSCIENCE2021, FerisePRL_2022}, PT-symmetry \cite{Bender1998, bender2007, feng2017non,NatPhysReviewEl-Ganainy2018, Ozdemir2019}, anti-PT-symmetry \cite{PengNAT2016,ChoiNATCOMM2018,YingliSCI2019,YangPhysRevLett.125.147202} and bound states in the continuum \cite{HsuNAT2016,YangPhysRevLett.125.147202,HanScienceAdv2023}.
It is known that coherent scattering of electromagnetic radiation is fully described by the positions of the poles and zeros of the scattering matrix \cite{Zanotto2014Nature,zanottoSCIREP2016,SweeneyPhysRevA.102.063511,HanScienceAdv2023,Rao2024NaturePhysics}.
In particular, for the reflection element of the scattering matrix (\textit{e.g.}, $S_{11}$), these are described by the eigenvalues of two slightly different, non-Hermitian effective Hamiltonians \cite{SweeneyPhysRevA.102.063511,HanScienceAdv2023,Rao2024NaturePhysics}. Hereafter, we will refer to them as $\hat H_{\rm res}$ and $\hat H_{\rm RZ}$, respectively, according to existing literature \cite{Rao2024NaturePhysics} (see \eqref{eq:S11_zeros} and \secref{sec_meth}). PA occurs when one or more eigenvalues of $\hat H_{\rm RZ}$ are on the real axis and the frequency $\omega$ of the input field matches such values. Ideal passive PT-symmetric systems display both the two relevant eigenvalues of $\hat H_{\rm RZ}$ on the real axis, thus enabling PA or CPA in the symmetry-unbroken phase. The full absorption of an input signal, which is often regarded as an undesired effect, can actually improve the performance of operations, such as detection \cite{WuPhysChemPhys2023, ChenNat2017}, quantum state transfer and wavelength conversion, and it may offer new opportunities in the development of optical modulators and switches or quantum sensing.

PA or CPA have been observed in optics using microdisks \cite{wangSCIENCE2021}, slab waveguides \cite{baranovNATREVMAT2017}, optical metamaterials \cite{rogerNATCOMM2015,fengPRB2012}, slabs of conductive materials \cite{muellersSCIADV2018} or nanostructured semiconductors \cite{Zanotto2014Nature}. These system are constituted by passive components and are fed with one or more input signals. The necessary balance between incoming energy and losses is achieved by adjusting the coupling of the system with its feeding lines and/or the losses of the system. PA can also be realized at microwave frequencies using metamaterials \cite{nieAPL2014}, conducting films \cite{liSCIREP2024,liPRB2015}, microwave resonators \cite{sunPRL2014,wennerPRL2014} or dielectrics \cite{chenNATCOMM2020}. Similar results have been obtained using microwave cavities coupled to ferromagnetic Yttrium-Iron-Garnet (YIG) spheres \cite{zhangPRB2023,zhangNATCOMM2017,qianNATCOMM2023}. Here, the system typically has a two ports configuration, which are used to send microwave excitations and, due to the low damping of YIG, its magnetic coupling with the cavity is adjusted to obtain the necessary critical coupling condition. 

Diluted spin centers in a non-magnetic matrix offer the possibility to play with multiple degrees of freedom, for instance, by choosing the crystal field or exploiting nuclear spins. These features can be tailored at the synthetic level in molecular spin systems \cite{santanniADVOPMAT2024,yamabayashiJACS2018,baderPhysChemChemPhys2017,baderNATCOMM2014} leading to some extensive control of their coherence times and over genuine quantum features \cite{candiniPRL2010,garlattiNATCOMM2017,timcoNATTECH2009}. These can be exploited for applications in quantum information processing \cite{ghirriMAGNETOCHEM2017,nakazawaAngChemIntern2012,chiesaJPHYSCHEMLETT2020,ranieriCHEMSCI2023} and in quantum sensing \cite{bonizzoniNPJQUANTINF2024,yuACSCENTSCI2021,troianiJMMM2019}. Finding optimal conditions for qubit encoding with molecular spins tipically requires tailoring the environmental bath/s \cite{atzoriJAMCHEMSOC2019} or being insensitive to it and/or to its fluctuations, as for \emph{atomic clock transitions} \cite{shiddiqNAT2016}. Conversely, running molecular spins as quantum sensors requires finding conditions for which small perturbations of the external parameters lead to large signal variations, \textit{e.g.} as those obtained at singularities in the scattering parameters. Finally, molecular spins can be embedded into hybrid spin-superconducting circuits \cite{bonizzoniADVPHYSX2018,bonizzoniNPJQUANT2020,chiesaPHYSREVAPPL2023,rollanoCOMMPHYS2022}, thus offering an ideal testbed to investigate open non-Hermitian systems at microwave frequency. 

\begin{figure*}[h!]
    \centering
    \includegraphics[width = 0.95\textwidth]{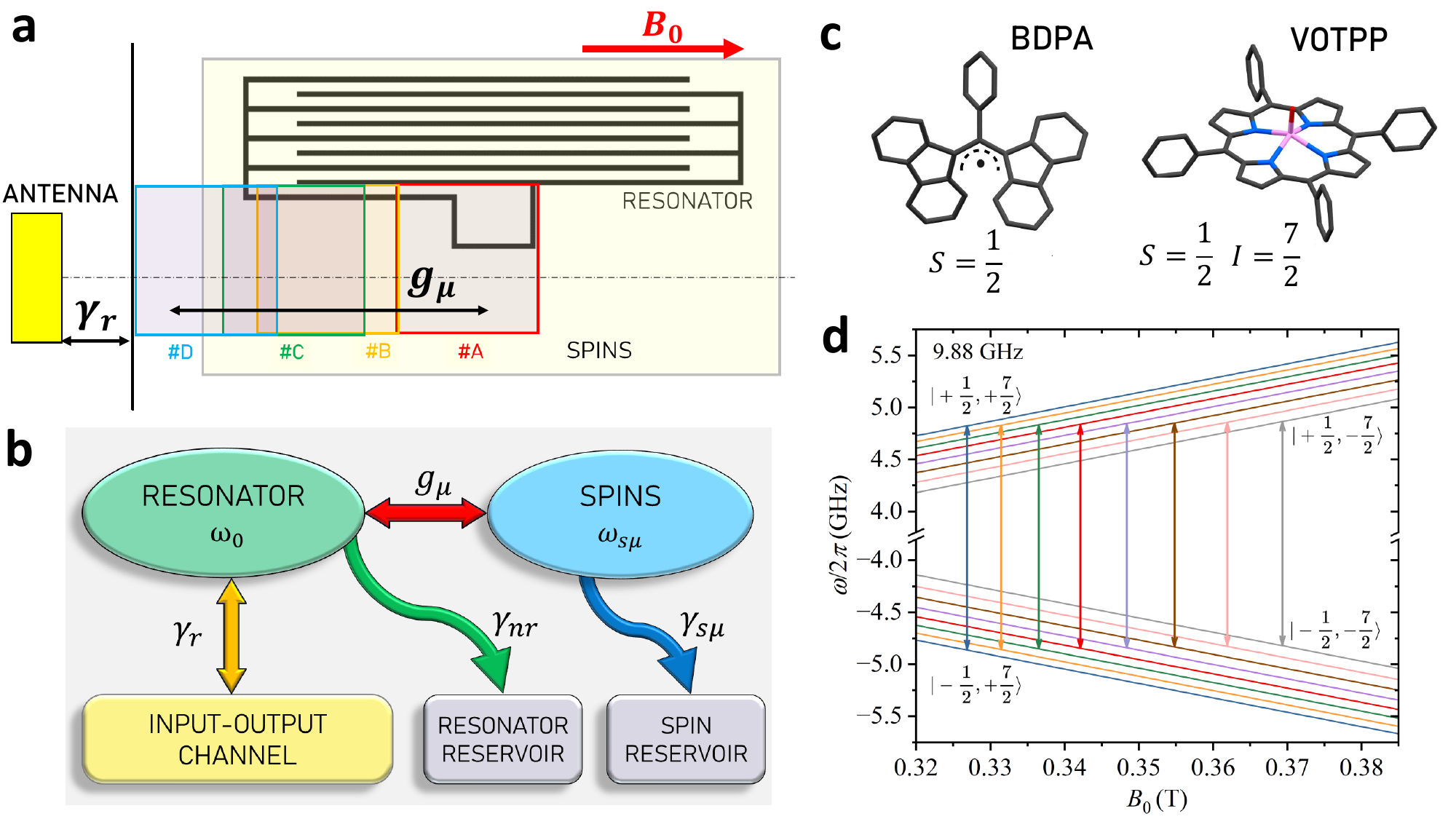} 
    \caption{\textbf{Implementing and modeling the open passive quantum system.}
    \textbf{a} Sketch of the lumped element resonator with all sample positions investigated in this work. The large light-yellow rectangle represents the BDPA sample, while the smaller rectangles represent the different positions of the VOTPP crystal (\#A to \#D, from red to light-blue). The distance between the antenna (yellow) and the chip can be adjusted at room temperature to vary the radiative relaxation rate of the resonator $\gamma_{r}$. The position of the sample controls the coupling strength $g_{\mu}$ with the resonator. The red arrow shows the direction of the applied static magnetic field, $B_{0}$. \textbf{b} Model adopted in this work for the open quantum system in \textbf{a}. The resonator is coupled to both spins and its input/output line (antenna). Only a single spin $\mu$-th ensemble is shown for clarity. \textbf{c} Molecular structure for BDPA and VOTPP. Labels indicate their electronic $S=1/2$ and nuclear $I=7/2$ spins. Images reproduced from \cite{yamabayashiJACS2018,Azuma2006} with permission. \textbf{d} Easyspin simulation of the $\{S_z,I_z\}$ energy levels of VOTPP obtained with the parameters and the Hamiltonian reported in \cite{yamabayashiJACS2018,bonizzoniNPJQUANT2020} at 9.88 GHz. Vertical arrows show the eight allowed $\ket{-\frac{1}{2},I_z} \leftrightarrow \ket{\frac{1}{2},I_z}$ transitions giving the $\mu$-th (sub)ensembles, denoted with different colors. The vertical energy scale is cut for better clarity.}
    \label{Fig1}
\end{figure*}

In this work, we theoretically and experimentally investigate an open passive system that is not PT-symmetric. Our geometry consists of a planar superconducting microwave resonator in the purely quantum regime (mK temperature and average single microwave photon number) overcoupled to its input-output line and magnetically coupled to molecular spin centers, as depicted in \figref{Fig1}. We first consider a diluted $\alpha,\gamma$-bisdiphenylene-$\beta$-phenylallyl (BDPA, for short) organic radical spin as a prototypical Two-Level-System (TLS) and, then, a tetraphenylporphyrinato oxovandium(IV) complex (VOTPP, for short), with large ($I=7/2$) nuclear spin and anisotropic hyperfine tensor to enlarge the number of subsystems investigated (\figref{Fig1} (d)). Tuning the position of the sample on the resonator allows us to change the coupling rate with the spins, $g_{\mu}$, thus crossing from the weak to the strong coupling regime. Moreover, the resulting non-equal thermal population of the hyperfine multiplet of VOTPP allows us to map the effect of slightly different couplings by simply tuning the applied static magnetic field. We experimentally demonstrate that it is possible to obtain PA with zero reflection dips at energy values detuned from resonance, at two symmetric positions with respect to it. On the one hand, this phenomenology provides a direct experimental confirmation consistent with theoretical predictions in Ref. \cite{zanottoSCIREP2016}. On the other hand, our detailed theoretical model and analysis allow us to directly link it to non-Hermitian physics. More specifically, we find that PA corresponds to tailor effective Hermitian subspaces in our open quantum system, even when radiative and non-radiative losses are not balanced, thus in the absence of PT-symmetry. This explicitly links PA with the concept of Hermitian subspaces and, at the same time, the presence of these subspaces largely shapes the overall aspect of the coherent spectra of light-matter systems as a function of detuning.

\section{Results}\label{sec_exp_res}
%

\subsection{Theoretical modeling}\label{sec_theory}

We begin by presenting the model developed for our system, which is sketched in \figref{Fig1}\,(a,b) (see \secref{sec_meth} for details). %
The resonator has frequency $\omega_0$, intrinsic (non-radiative) losses, $\gamma_{\rm nr}$, and it is coupled to an antenna acting as an input-output port with rate $\gamma_{\rm r}$. The $\mu$-{th} spin ensemble is coupled to the resonator through its collective coupling strength $g_\mu$, and has energy $\omega_{{\rm s} \mu} = g_{L\mu} \mu_B B_0 / \hbar$ (being $B_0$ the applied static magnetic field, $g_L$ is an effective Landé g-factor and $\mu_B$ is the Bohr's magneton) and intrinsic relaxation rate $\gamma_{{\rm s} \mu}$. While BDPA can be effectively described using a single resonance frequency ($\mu=1$), the VOTPP sample comprises multiple molecular spin levels due to hyperfine splitting ($\mu=1,\dots ,8$), in \figref{Fig1}\,(d). These multiple-spin ensembles can be effectively described using the Holstein-Primakoff mapping with $N$ non-interacting bosons. In the low-excitation regime, such mapping can be linearized, leading to the harmonic Hamiltonian
\begin{equation} \label{eq:H_S}
    \hat{H}_{S} = \hbar \omega_0\, \hat{a}^\dagger \hat{a} + \hbar \sum_{\mu=1}^N \omega_{s\mu}\, \hat{b}_\mu^\dagger \hat{b}_\mu + \hbar \sum_{\mu=1}^N g_\mu(\hat{a}^\dagger \hat{b}_\mu+ \hat{b}_\mu^\dagger \hat{a})\,,  
\end{equation}
where $\hat{a}$ and $\hat{b}_\mu$ denote the bosonic annihilation operators of the resonator and the $\mu$-th spin ensemble, respectively. The coupling strengths $g_\mu$ observed are sufficiently small with respect to $\omega_0$ to justify the use of the Rotating Wave Approximation (RWA) \cite{FriskKockum2019}. \eqref{eq:H_S} can describe either BDPA for $N=1$ or VOTPP for $N=8$.
The coupling with the external environment, which consists of the decay channels of the resonator (radiative and non-radiative) and of the spin ensembles (\figref{Fig1} (b)), can be conveniently described by Heisenberg-Langevin equations, which take into account also the coherent feeding field through the antenna (see Supplementary Information). The complex reflection scattering parameter, $S_{11}(\omega)$, can be shown to have poles (\textit{i.e.}, resonance states) corresponding to the complex eigenfrequencies, $\Omega_j$, of the non-Hermitian Hamiltonian $\hat{H}_{\rm res} / \hbar = \hat{\boldsymbol{\alpha}}^\dagger (\mathbf{A} - i \mathbf{\Gamma}/2) \hat{\boldsymbol{\alpha}}$, where $\hat{\boldsymbol{\alpha}}^{\rm T} = (\hat{a}, \hat{\mathbf{b}})$, $\mathbf{A}$ is the Hopfield matrix of Hamiltonian in \eqref{eq:H_S} and $\mathbf{\Gamma}$ is the corresponding decay matrix.
Moreover, we observe that the zeros of $S_{11}(\omega)$ exhibit the analogous structure as the denominator, with the only difference being the reversal of the sign of the decay rate associated with the input-output port, $\gamma_{\rm r}$ (see \secref{sec_meth}) \cite{zanottoSCIREP2016,SweeneyPhysRevA.102.063511,lamberto2024superradiantquantumphasetransition,Rao2024NaturePhysics}. Therefore, these reflection zeros can be interpreted as the complex eigenfrequencies, $\tilde{\Omega}_j$, of the effective non-Hermitian Hamiltonian
\begin{equation} \label{eq:H_eff}
    \hat{H}_{\rm RZ}/\hbar = \hat{\boldsymbol{\alpha}}^\dagger (\mathbf{A} - i \tilde{\mathbf{\Gamma}}/2) \hat{\boldsymbol{\alpha}} \, ,
\end{equation}
where the decay matrix $\tilde{\mathbf{\Gamma}}$ differs from $\mathbf{\Gamma}$ only for the inversion of the sign of the input-output port decay rate (notice that $-\gamma_r$ corresponds to the feeding rate). Hence, the reflection scattering parameter reads:
\begin{equation} \label{eq:S11_zeros}
    S_{11}(\omega) = \frac{\det \left( \hat{H}_{\rm RZ} - \omega \hat{I}\right)}{\det \left( \hat{H}_{\rm res} - \omega \hat{I}\right)} = \prod_{j=1}^{N+1} \frac{\omega-\tilde \Omega_j}{\omega-\Omega_j} \, ,
\end{equation}
where $\hat{I}$ is the identity operator. Here we notice that, in \eqref{eq:S11_zeros} we have implicitly performed an analytic continuation of the reflection scattering parameter $S_{11}(\omega)$ into the complex $\omega$-plane. However, in the context of our experimental implementation, $\omega$ must be real-valued.

Equation (\ref{eq:S11_zeros}) explicitly highlights the connection between the PA condition ($S_{11}(\omega) = 0$) and the eigenvalues of the effective non-Hermitian Hamiltonian $\hat{H}_{\rm RZ}$ in \eqref{eq:H_eff}. Indeed, when the imaginary part of one of the complex eigenfrequencies of \eqref{eq:H_eff} vanishes ($\Im(\tilde{\Omega}_j)=0$ for some $j$), it is always possible to nullify the numerator of $S_{11}$ by appropriately tuning the frequency of the input probing field $\omega$. Reflection zeros on the real axis are also referred to as Reflectionless Scattering Modes (RSM) \cite{SweeneyPhysRevA.102.063511,Rao2024NaturePhysics}.
To further elucidate this point, we first perform a rotation on the bare bosonic operators to transform them in the polariton basis (normal modes). Specifically, we introduce a suitable Hopfield transformation $\bf U$, defined as $\hat{\mathbf{P}} = \mathbf{U} \, \hat{\boldsymbol{\alpha}}$, such that the Hamiltonian of the isolated system in \eqref{eq:H_S} takes the diagonal form $\hat{H}_S = \sum_{j=1}^{N+1} \bar{\omega}_j \hat P^\dagger_j \hat P_j$. Subsequently, the effective non-Hermitian Hamiltonian describing the open system dynamics of the reflection zeros, in the strong coupling regime, is given by
\begin{equation} \label{eq:H_eff_polariton}
    \hat{H}_{\rm RZ} = \sum_{j=1}^{N+1}\left(\bar{\omega}_j - i \frac{\bar{\gamma}_j}{2}\right) \hat{P}_j^\dagger \hat{P}_j \, ,
\end{equation}
where the effective (or \emph{dressed}) loss rates $\bar{\gamma}_j$ are defined by the diagonal elements of ${\bf U} \tilde{\bf \Gamma} {\bf U}^\dagger$ (see Methods).
(\ref{eq:H_eff_polariton}) provides a clear physical interpretation of the reflection zeros in terms of the polariton modes. In particular, the PA condition turns out to be $\bar{\gamma}_j=0$ for some $j$. We remark that this effective Hamiltonian is derived under the hypothesis of PA and is valid in the strong coupling regime, as, in this regime, the mixed terms in the decay channels can be safely neglected due to the different resonance frequencies of the polariton peaks.

\subsection{The Single Ensemble Case}\label{sec_single}

\begin{figure*}[h!]
    \centering
    \includegraphics[width =0.95 \textwidth]{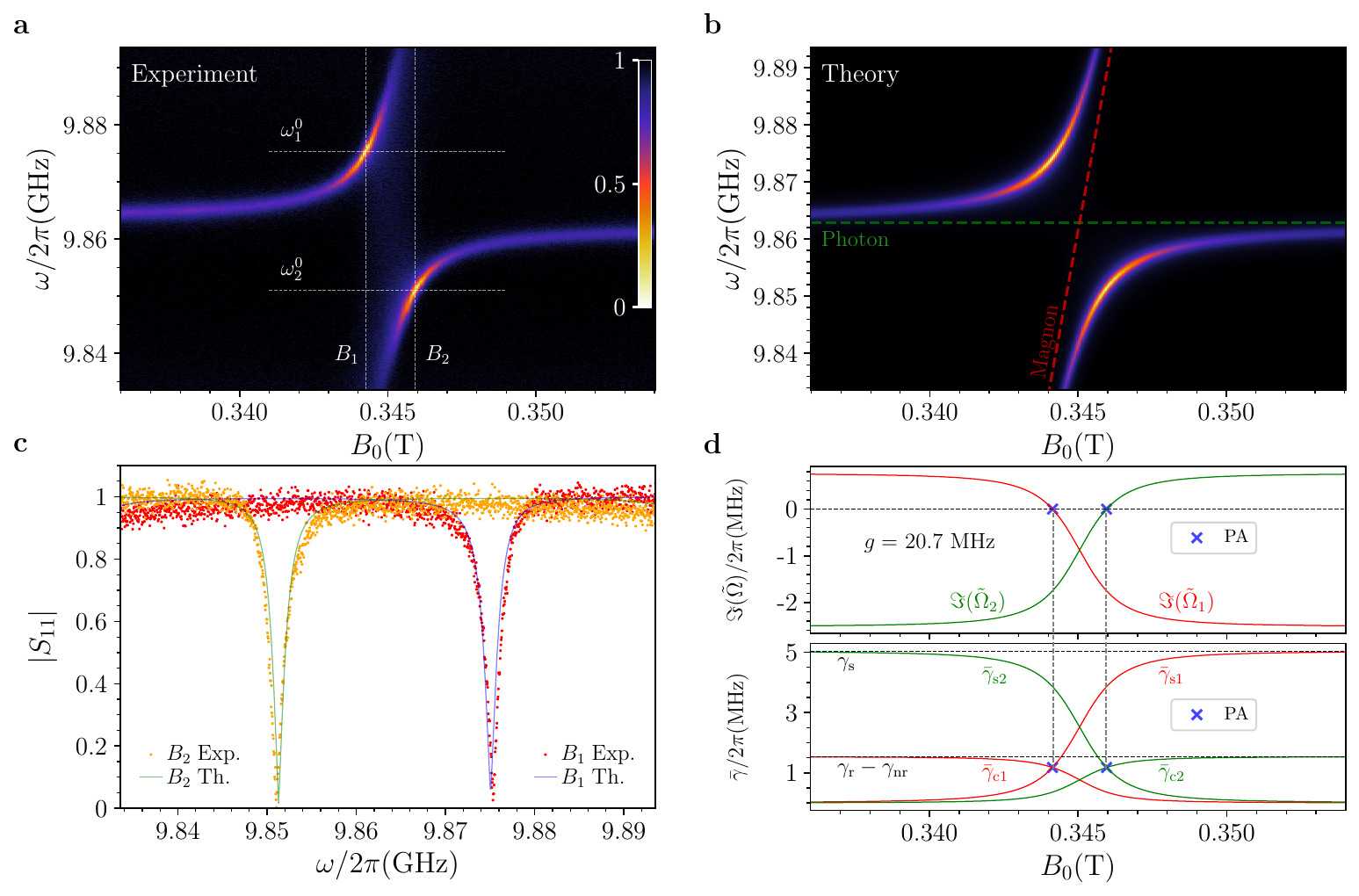}		
    \caption{\textbf{Perfect Absorption for the Single Ensemble Case.}  
    \textbf{a} Normalized reflection map (\(|S_{11}(\omega)|\)) measured for the BDPA sample at 25 mK as a function of the static magnetic field $B_0$.  
    \textbf{b} Simulated reflection map obtained using the fit parameters extracted from the map in panel \textbf{a}, according to \eqref{eq:meth-S11} in Methods (fit parameters are reported in Table~1 of the Supplementary Information). 
    \textbf{c} Theoretical (green and blue lines) and experimental (orange and red dots) normalized reflection spectra extracted from the maps in \textbf{a} and \textbf{b}, showing two dips with nearly zero reflection (see vertical lines $B_1$ and $B_2$ in panel \textbf{a}).  
    \textbf{d} (Upper panel) Imaginary parts of $\tilde{ \Omega}_{1,2}$, $[\Im(\tilde{\Omega}_{1,2})]$, as a function of $B_0$, calculated using \eqref{eq:S11_zeros} and the fit parameters obtained from \text{a}. Perfect absorption (blue crosses) occurs at the polariton resonances when \(\Im(\tilde{\Omega}_{1,2})\) crosses zero.  
    \textbf{d} (Lower panel) Dressed cavity feeding rate $\bar\gamma_{{\rm c}i}$ and dressed spin loss rate $\bar\gamma_{{\rm s}i}$ as functions of $B_0$, computed according to \eqref{eq:gamma_bar_strong}. Perfect absorption (blue crosses) is achieved when $\bar\gamma_{{\rm c}i}=\bar\gamma_{{\rm s}i}$ for the $i$-th polariton (see main text). 
    }
    \label{fig:single_spin}
\end{figure*}

We begin by analyzing the BDPA organic radical\figref{fig:single_spin} (a) displays experimental reflection spectra maps as a function of the static magnetic field. \figref{fig:single_spin} (b) illustrates the corresponding theoretical fit performed using \eqref{eq:S11_zeros} with $N=1$, showing excellent agreement.
An anticrossing, which is typical of the strong spin-photon coupling regime, is clearly visible when $\omega_{\rm s}$ crosses $\omega_0$. Fits gives $g/2 \pi \approx 20.7 $ MHz and $\gamma_{\rm s}/2\pi = 5$ MHz, further corroborating the strong coupling regime (see Table S1 of Supplementary Information for full fit parameters). 
We observe the presence of two dips (\figref{fig:single_spin} (c)), approaching zero reflection, corresponding to PA.
Remarkably, these occur before ($0.3443$ T) and after ($0.3459$ T) the resonant magnetic field value, corresponding to symmetric resonator-spin detuning $\Delta = \omega_{\rm s} - \omega_0 \approx  \pm 8 \cdot 10^{-4}$ T. This clearly differs from what is typically observed in PT-symmetric systems, where the dips occur at resonance \cite{zhangNATCOMM2017, wangSCIENCE2021}. The positions and the absolute values of the two reflection zeros are also predicted by our model (\figref{fig:single_spin} (b)).  As a first important result, this demonstrates (both experimentally and theoretically) that PA can be realized on molecular spin centers by tuning their resonance frequency. More specifically, in the strong coupling regime, the imaginary part of the $j$-th effective complex eigenfrequencies, $\Im(\tilde{\Omega}_j)$, is approximately equal to the effective loss rates $\bar{\gamma}_j / 2$, as given by \eqref{eq:H_eff_polariton}, which, in turn, can be expressed by a linear combination of the single-channel loss rates (see Methods):

\begin{equation} \label{eq:gamma_bar_strong}
    \bar{\gamma}_j = (-\gamma_{\rm r}+\gamma_{\rm nr}) \, \abs{U_{j1}}^2 + \gamma_{\rm s} \, \abs{U_{j2}}^2 = -\bar{\gamma}_{{\rm c} j} + \bar{\gamma}_{{\rm s} j} \, .
\end{equation}
Here, $U_{jk}$ are the Hopfield coefficients defined by $\bf U$, which determines the degree of photon ($\abs{U_{j1}}^2$) and spin ($\abs{U_{j2}}^2$) hybridization of the polaritons, and which can be tuned through the spin-resonator detuning. $\bar{\gamma}_{{\rm c} j} = (\gamma_{\rm r} - \gamma_{\rm nr}) \, \abs{U_{j1}}^2$ and $\bar{\gamma}_{{\rm s} j} = \gamma_{\rm s} \, \abs{U_{j2}}^2$ represent the \emph{dressed} cavity feeding rate and the \emph{dressed} spin loss rate, respectively. In our experiments, we vary the detuning, $\Delta$, through the externally-applied magnetic field.
Therefore, PA (defined by $\bar{\gamma}_j = 0$) occurs when the balance between \emph{dressed} feeding and loss rates is achieved, analogous to the balance condition in PT-symmetric systems for the \emph{bare} rates, even if the Hamiltonian does not display such symmetry (see \secref{sec_disc} for a more detailed comparison with PT-symmetric systems).
Equation (\ref{eq:gamma_bar_strong}) can be satisfied for either the lower or upper polariton, leading to the emergence of a Hermitian subspace within the Hilbert space of $\hat{H}_{\rm RZ}$. This subspace $\mathcal{H}^{(j)}$ is effectively described by the Hamiltonian $\hat{H}_{\rm HS}^{(j)} = \bar{\omega}_j \hat{P}_j^\dagger \hat{P}_j$, where $j=1$ or $j=2$ for the upper or lower polariton, respectively. Furthermore, it can be easily shown that if a Hermitian subspace exists for one of the polariton modes at a given detuning $\Delta'$, a second one will exist at $-\Delta'$ for the other polariton mode. This property holds even in the case of multiple spin resonances, as it will be shown for the VOTPP sample. \figref{fig:single_spin} (d) shows the imaginary part of the complex eigenvalues of \eqref{eq:H_eff}, $\Im(\tilde{\Omega}_j)$, along with the corresponding contributions to the $j$-th polariton mode of the dressed cavity feeding and spin loss rates. The balance between the two contributions ($\bar{\gamma}_{cj} = \bar{\gamma}_{sj}$), corresponding to the simultaneous vanishing of $\Im(\tilde{\Omega}_j)$, is observed for either of the polariton modes at symmetric detunings, as discussed above.

\subsection{Effects of coupling strength}\label{sec_multi}

\begin{figure*}[t]
    \centering
    \includegraphics[width =0.9\textwidth]{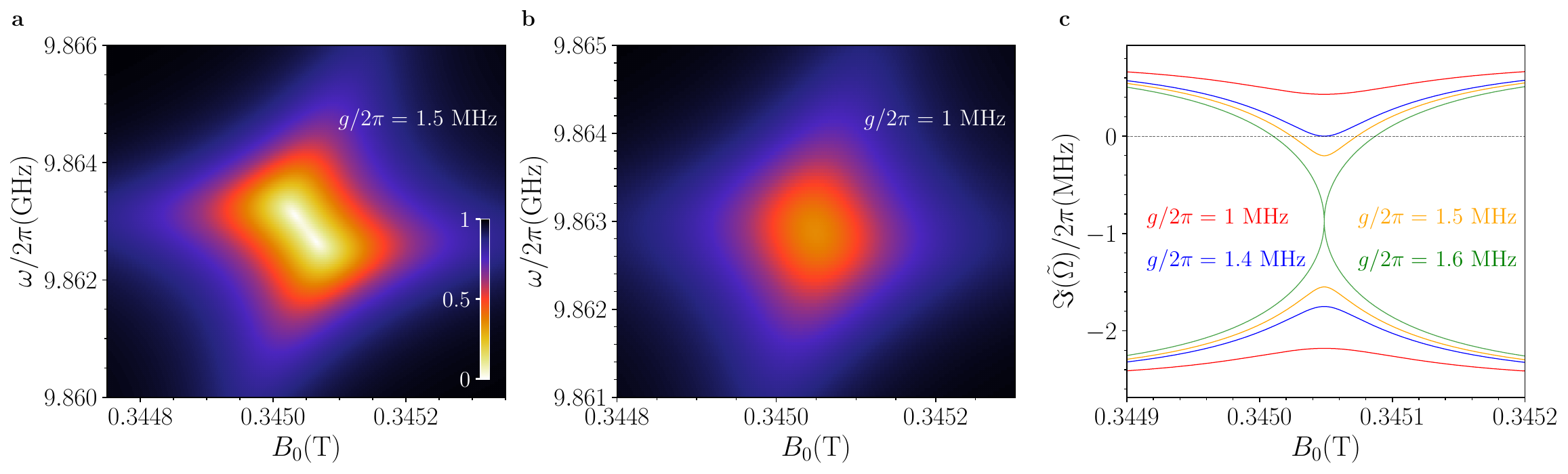}
    \caption{\textbf{Perfect Absorption During the Transition from Strong to Weak Coupling Regime.}  
    \textbf{a,\,b} Normalized reflection ($\abs{S_{11}}$) maps as a function of the static magnetic field $B_0$, simulated using \eqref{eq:meth-S11} in Methods with the parameters fitted from the data in \figref{fig:single_spin}, except for using lower coupling strength values of $g/2\pi = 1.5\,$MHz (\textbf{a}) and $g/2\pi = 1\,$MHz (\textbf{b}), respectively. \textbf{c} Imaginary parts of $\tilde{ \Omega}_{1,2}$, $[\Im(\tilde{\Omega}_{1,2})]$, calculated as a function of the static magnetic field using the relaxation rates fitted from the data in \figref{fig:single_spin} and four different values of $g/2\pi$. Perfect absorption occurs when \(\Im(\tilde{\Omega}_{1,2})\) crosses zero and cannot be realized for $g/2\pi <\,1.4\,$MHz (see main text).}
    \label{fig:single_spin_sim}
\end{figure*}

We now investigate theoretically the effect of the coupling strength on PA. \figref{fig:single_spin_sim} shows reflection maps simulated with \eqref{eq:meth-S11} by using the relaxation rates obtained from the fits of \figref{fig:single_spin}, along with different coupling strengths. Notably, as the coupling decreases the dips progressively merge towards the resonant field value. We identify a threshold value, $g_{\rm th} = (\gamma_{\rm r} - \gamma_{\rm nr} + \gamma_{\rm s}) / 4$ (green line in \figref{fig:single_spin_sim} (c)), marking the transition from the strong to the weak coupling regime.
An intermediate region emerges, where no crossing of the imaginary parts occurs anymore but PA still persists (yellow line), until the coupling strength reaches a minimum value $g_{\rm min} = \sqrt{(\gamma_{\rm r} - \gamma_{\rm nr}) \gamma_{\rm s}} / 2$, at which $\Im(\tilde{\Omega}_j)$ exhibits a double root at zero detuning (blue line) and the reflection dips coalesce into a single one. For couplings below $g_{\rm min}$, the imaginary part $\Im(\tilde{\Omega})$ cannot cross the zero, preventing the observation of PA (red line). The simulations in \figref{fig:single_spin_sim} suggest that our model in \secref{sec_theory}, although developed under the assumption of strong coupling regime, holds also in the weak coupling regime.

\subsection{The Multiple Ensemble Case}\label{sec_multi}

We now experimentally investigate an analogous phenomenology using the VOTPP sample, which, in contrast to BDPA, exhibits multiple resonance frequencies $\omega_{{\rm s}\mu}$. As mentioned in the introduction, the overall coupling regime for the whole group of transitions is primarily determined by the position of the sample with respect to the resonator (see \figref{Fig1}). In addition, the values of the hyperfine tensor of VOTPP (see \secref{sec_meth}) give different thermal population of hyperfine levels at 30 mK, thus allowing us to span a large set of $g_j$ values within a unique field scan. 
%
\begin{figure*}[h!]
    \centering
    \includegraphics[width =0.9 \textwidth]{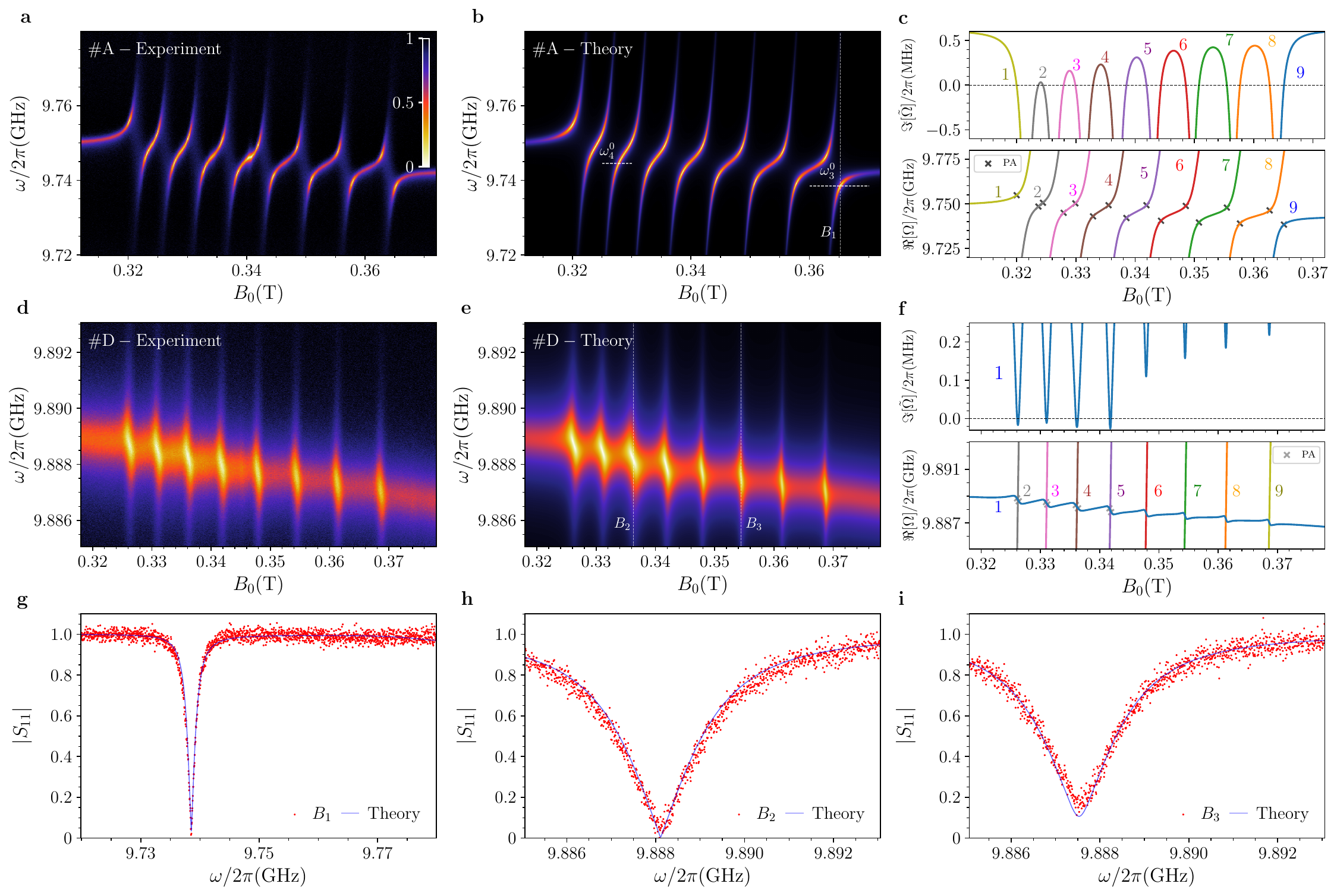}
    \caption{\textbf{Perfect Absorption for the Multiple Spin Case. } \textbf{a,\,d} Normalized reflection ($\abs{S_{11}}$) maps measured as a function of the static magnetic field $B_0$ at 30 mK for the VOTPP crystal. PA is observed in proximity of multiple hyperfine levels. The strongest (position $\#A$) and the weakest (position $\#D$) coupling regimes are shown, respectively. \textbf{b,\,e}
    Simulated reflection maps obtained by fitting the maps in \textbf{a},\textbf{d} according to \eqref{eq:meth-S11} in Methods.
    \textbf{c,\,f} (Upper panels) Imaginary parts of the complex frequencies, $\Im{(\tilde{\Omega}_j)}$, as a function of the magnetic field $B_0$, for the $j$-th polariton frequency. The horizontal black dotted line at $\Im{(\tilde{\Omega})}=0$ corresponds to the PA condition. \textbf{c,\,f} (Lower panels) Real parts of the polariton frequencies, $\Re{(\Omega_j)}$, with the predicted PA points, showing excellent agreement with the experimental data. Both $\Im{(\tilde{\Omega}_j)}$ and $\Re{(\Omega_j)}$ are calculated using \eqref{eq:S11_zeros} supported by \eqref{eq:meth-S11} in Methods. \textbf{g,\,h, \,i} Experimental (red dots) and theoretical (blue lines) normalized reflection obtained according to the vertical lines shown in  \textbf{b,\,e} ($B_1$, $B_2$, and $B_3$), displaying perfect absorption dips and one not satisfying this condition. All fit parameters are given in tables~5 and 8 of Supplementary Information. 
}
\label{fig:multi_spin}
\end{figure*}

Figure ~\ref{fig:multi_spin} (a,\,d) shows experimental reflection maps, $\abs{S_{11}}$, measured as function of the static magnetic field, for positions $\#A$ and $\#D$ of VOTPP (see \figref{Fig1}). In position $\#A$, the observed avoided level crossings clearly indicate that the high cooperativity regime is achieved on each line. This is also supported by the fitted values $g_{\mu}/2\pi=13-16\,$ MHz and $\gamma_{{s} \mu}/2\pi=7-13\,$ MHz (depending on the line, see Table S5 of Supplementary Information). A pair of dips in reflection, approaching near-zero values and at non-zero symmetric detunings from resonance, is clearly visible for each line, indicating the presence of PA. Conversely, data for position $\#D$ show the weak coupling regime for all resonances and significantly lower coupling strengths (see Table S8 of Supplementary Information). Here, due to the different thermal population of each line, PA occurs only for a subset of the eight resonant lines (from left to right, the first four), and the coalescence of those dips is observed for increasing magnetic field values. The theoretical reflection maps in \figref{fig:multi_spin} (b,e), simulated through \eqref{eq:S11_zeros}, are in excellent agreement with the data. For position \#A, the PA points are perfectly predicted by the zeros of the imaginary parts of the corresponding effective eigenfrequencies (each line crosses the zero, see \figref{fig:multi_spin} (c)), which are calculated by using the fit parameters obtained from \figref{fig:multi_spin} (a). Conversely, for position \#D, the imaginary parts of the complex eigenvalues cross zero only for the first four lines. This can be explained by the minimal coupling $g_{\rm min}$ discussed in \secref{sec_single}, since moving across the different resonances results in the coupling strengths $g_j$ becoming smaller than $g_{\rm min}$.
These latter results further demonstrate that \eqref{eq:S11_zeros} holds for any coupling regime, demonstrating that the PA condition can be achieved provided that a minimum coupling rate value is overcome for a given set of relaxation rates.

\section{Discussion}\label{sec_disc}

We have experimentally realized PA into a passive open quantum system composed of molecular spin centers, both in the strong and the weak coupling regimes with a planar superconducting microwave resonator.  Our results show that spins and, more specifically, molecular spins, turn out to be an excellent testbed for investigating non-Hermitian physics, thanks to the possibility of tuning different parameters and spanning over different coupling regimes.

We observe that our experimental results refer to a regime for which the system is not PT-symmetric, and that, in principle, PT-symmetry could be realized in these platforms by further tuning system parameters (\textit{e.g.}, by modifying the position of the antenna relative to the resonator, and properly tuning the static magnetic field). 
However, we note that our implementation is less constrained than PT-symmetric configurations, as light-matter detuning can be easily adjusted to achieve PA without strictly satisfying the loss balance condition required by PA in PT-symmetric systems. In this regard, we provided a simple yet physically insightful interpretation linking the imaginary part of the reflection zeros to the spin and photon content of the polaritons, through the Hopfield coefficients. These coefficients (and thus the position of the reflection zeros on the complex plane) can be dynamically tuned by varying the resonance frequency of one of the subsystems (in our work, by acting on the external magnetic field), which enables the movement of the reflection zeros on and off the real axis and the realization of a Hermitian subspace. This flexibility makes our platform promising not only for exploring PA in coherently coupled systems, but also for studying a broader class of non-Hermitian phenomena in passive open quantum systems. These include dissipative couplings mediated by waveguides \cite{Rao2024NaturePhysics,wangSCIENCE2021}, as well as non-Hermitian topological effects \cite{TangPhysRevLett2021,WangNature2022}.

\begin{figure*}[h!]
    \centering
    \includegraphics[width=0.85\linewidth]{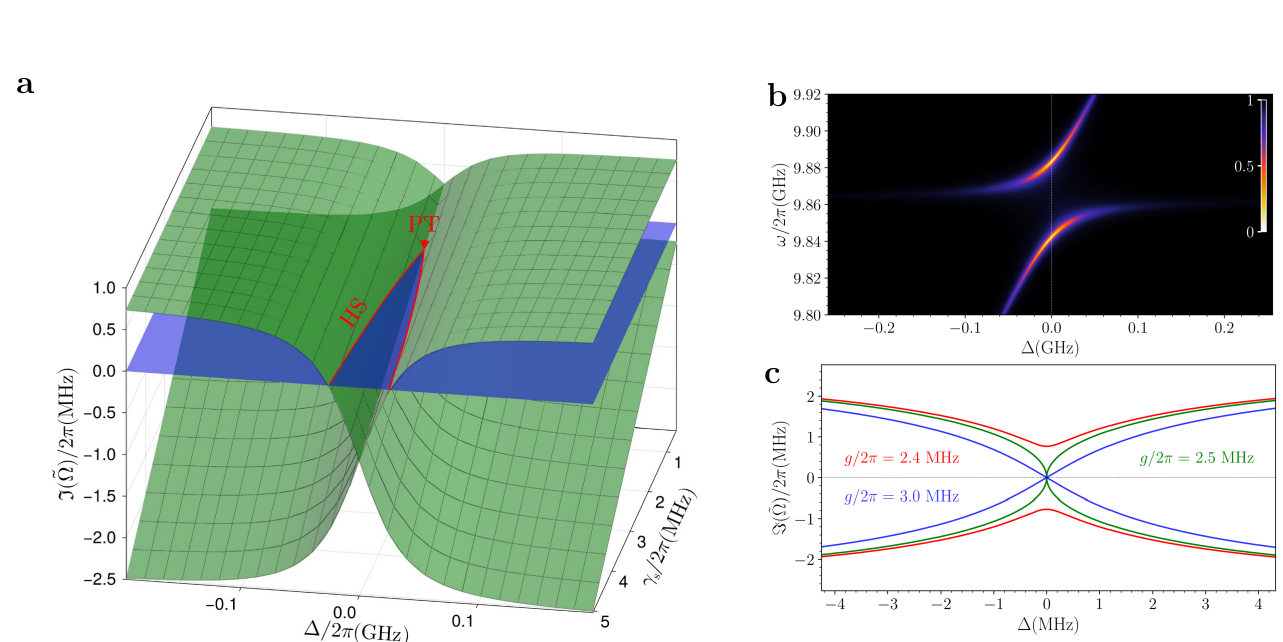}
    \caption{\textbf{Comparison with PT symmetry.}  
    \textbf{ a} Surface plot showing the imaginary part of the complex eigenvalues of the effective non-Hermitian Hamiltonian $\hat{H}_{\rm RZ}$ as a function fo the detuning $\Delta$ and the spins decay rate $\gamma_{\rm s}$. The simulation is obtained using $N=1$ and the values reported in Table S1. The two red lines indicate where an Hermitian subspace is realized, i.e., the intersection with the $\Im (\tilde{\Omega}) = 0$ plane (in blue), where one of the eigenvalues becomes real. These lines coalesce at the point $(\gamma_{\rm s}, \Delta) = (\gamma_{\rm r} - \gamma_{\rm nr}, 0)$, marked by the red triangle, corresponding to the PT symmetry condition. 
    \textbf{b} Normalized reflection map as a function of the detuning for a PT-symmetric system in the strong coupling regime. PA is achieved simultaneously for both polariton branches at $\Delta = 0$, conversely to the result presented in this work. 
    \textbf{c} Imaginary parts of the eigenfrequencies $\Im (\tilde{\Omega})$ as a function of the detuning for different coupling strengths, ranging from the weak to the strong coupling regime, in a PT-symmetric system. } 
    \label{fig:3d_plot}
\end{figure*}

Our theoretical framework is consistent with previous findings observed in PT-symmetric systems (\textit{e.g.}, Ref. \cite{zhangNATCOMM2017,wangSCIENCE2021}), of which it can be regarded as a generalization. More specifically, when the system is on resonance, $\omega_0 = \omega_{\rm s}$ (i.e. $\Delta = 0$), and the decay rates satisfy $\gamma_{\rm r} - \gamma_{\rm nr} = \gamma_{\rm s} \equiv \gamma$, the effective non-Hermitian Hamiltonian describing the reflection zeros, $\hat{H}_{\rm RZ}$, becomes PT-symmetric (\figref{fig:3d_plot} (a)). Under these conditions, the Hopfield coefficients satisfy $\abs{U_{j1}}^2 = \abs{U_{j2}}^2 = 1/2$, and the balance condition in \eqref{eq:gamma_bar_strong} is
satisfied for both the polariton modes. This behavior is clearly confirmed in the reflection spectra, where both dips reach zero for $\Delta = 0$ (\figref{fig:3d_plot} (b)).
As the coupling strength is decreased, the system reaches the Exceptional Point at $g_{\rm EP} = \gamma/2$ (green line in \figref{fig:3d_plot} (c)). Given that in PT-symmetric systems the relation $g_{\rm th} = g_{\rm min} = g_{\rm EP}$ holds, no intermediate regime appears as the coupling decreases and, thus, the presence of a crossing in the imaginary parts of the eigenvalues, $\Im(\tilde{\Omega}_i)$, can be directly associated with the presence of PA. Indeed, in PT-symmetric systems $\Im(\tilde{\Omega}_i)$ is symmetric with respect to the real axis and exhibits a double zero (at $\Delta = 0$) only for $g>g_{\rm EP}$. 
Furthermore, our theoretical framework holds even in the presence of multiple resonances, as the ones of VOTPP. This goes beyond traditional PT-symmetric models, as the notion of PT symmetry becomes ill-defined in systems with more than two coupled subsystems.\\

\begin{figure*}[h]
    \centering
    \includegraphics[width=\linewidth]{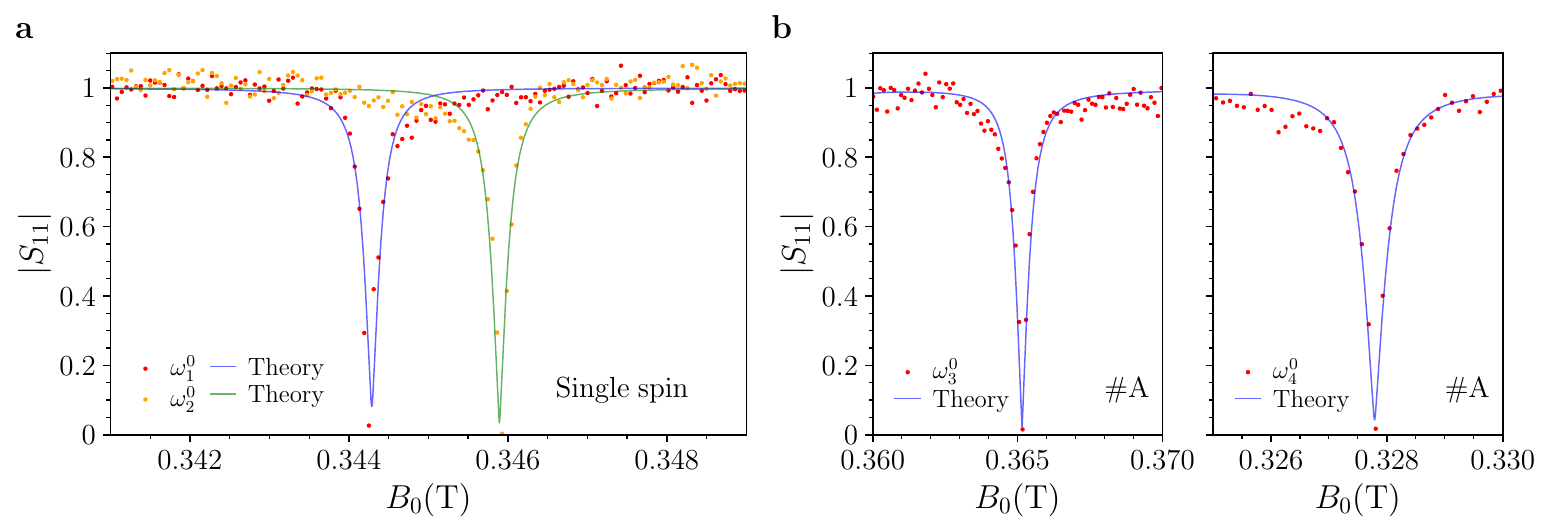}
    \caption{\textbf{Potential Implementation of Single Microwave Photon Switches.}  
    \textbf{a} Experimental (red and orange dots) and theoretical (blue and green lines) normalized reflection ($|S_{11}|$) as a function of the static magnetic field $B_0$ for the BDPA sample (single spin case), highlighting perfect absorption. Data are extracted along the horizontal lines shown in \figref{fig:single_spin}(a) (at fixed frequency, $\omega^0_1$ and $\omega^0_2$).  
    \textbf{b} Experimental (red dots) and theoretical (blue lines) normalized reflection $|S_{11}|$ obtained for the VOTPP crystal (position \#A) under PA and} based on the horizontal lines illustrated in \figref{fig:multi_spin}(b) (see $\omega_3^0$ and $\omega^0_4$).    
    \label{fig_modulation}
\end{figure*}

Our experiments are carried out in the purely quantum regime, which is milliKelvin temperature and average single microwave photon number in the resonator. Although PA is a general phenomenon of resonant systems, including classical ones, the model we use is perfectly suitable to describe quantum systems. For instance, this is relevant in the view of exploring how non-Hermitian physics can affect non-classical effects (see, \textit{e.g.}, Ref. \cite{wang2024}). 

As an example for applications, we consider fast single-photon switches or modulators for microwave radiation \cite{feng2017non}. \figref{fig_modulation} displays reflection measurements ($|S_{11}|$) and corresponding theoretical fits, as a function of the externally applied magnetic field $B_0$ (\text{i.e.}, at a fixed microwave frequency), corresponding to the horizontal line-cuts shown in \figref{fig:single_spin} (a) and \figref{fig:multi_spin} (b,e). A rather small variation of the magnetic field ($\sim 10^{-4}$ T) can switch the reflectivity from its maximum value to nearly zero. In this way, the reflection can be switched suppressing the single microwave photon or allowing it to propagate by simply applying a time-dependent local magnetic field bias (\textit{e.g.}, through a modulation coil). We can estimate the achievable modulation depth from the data in \figref{fig_modulation} as: 
\begin{equation}
    M_d = 20\, \log_{10}\left[ \frac{|S_{11}^{\rm on}|}{|S_{11}^{\rm off}|}\right] \, ,
\end{equation}
where $|S_{11}^{\rm on (off)}|$ represents the maximum (minimum) value of $\abs{S_{11}}$.
Without specific optimization of the system, we find $M_d \approx 50\,\rm dB$ for the BDPA by varying $B_0$ within a range of approximately $4\cdot10^{-4}\,\rm T$ (\figref{fig_modulation} (a)). Notably, this value remains robust against fluctuations in the \textit{on}-state (maximum $S_{11}$), as the near-zero reflectivity in the \textit{off}-state (minimum $S_{11}$) dominates the modulation depth. For the VOTPP sample in position $\#A$ (\figref{fig_modulation} (b)) we observe a shallower dip compared to BDPA. However, we obtain $M_d \approx 35\,\rm dB$ using lines $\mu=3,4$. 

Further applications can be foreseen by noticing that singularities in spectra are of interest for sensing \cite{wiersigPHOTRES2020,degenREVMODPHYS2017,goryachevPHYSDARKUNIV2019}. Here, a possible detection mechanism relies on the variation of the reflection signal ($|S_{11}|$) upon the strength of an added perturbation (\emph{slope detection} \cite{degenREVMODPHYS2017}). For instance, based on the data in \figref{fig_modulation}.a,b, and without any specific optimization of the system, we can estimate a slope of $|\frac{\Delta S_{11}}{\Delta B_0}| \approx 2016\, T^{-1}$, corresponding to a transduction coefficient of $\approx 5\cdot10^{-4} T$ for unit of reflection around the PA point. Here, a potential advantage of PA relies on its relatively simple phenomenology and on the reduced intrinsic noise and fluctuations, which, otherwise, could be a limiting factor for sensing \cite{wiersigPHOTRES2020}. Detection of electromagnetic radiation could, in principle, benefit as well from our results after proper extension and optimization of our system (\textit{e.g.} adding an additional input transmission line or antenna to route the incoming radiation to the sensor or using photoresponsive molecular spins \cite{santanniADVOPMAT2024}). For instance, detection of itinerant single microwave photons would help in searching for rare events, such as Dark Matter Axions \cite{lasenbyPRD2021,goryachevPHYSDARKUNIV2019}.

We finally mention that our results can be readily transferred and applied also on different paramagnetic spin centers, including defects such as Er$^{3+}$ ions, P donors in Si and Nitrogen-Vacancy centers and, potentially, further extended and applied to very different frequency ranges and platforms (\textit{e.g.}, optical frequency).

\section{Methods}\label{sec_meth}

\subsection{Derivation of the Reflection Scattering Parameter}\label{subsec_theor}

To derive the reflection spectrum, we first express the output fields in terms of the input fields using the quantum Langevin equations. This leads to the general relation (see Supplemental Material)
\begin{equation} \label{eq:meth-output_vs_input}
    \hat{\mathbf{F}}_{\rm out}(\omega) = \left[-\frac{1}{2}\mathbf{\Gamma}-i(\omega \,\mathbf{I}-{\bf A})\right]\left[\frac{1}{2}\mathbf{\Gamma}-i(\omega \,\mathbf{I}-{\bf A})\right]^{-1} \, \hat{\mathbf{F}}_{\rm in}(\omega) \,\, ,
\end{equation}
where $\hat{\mathbf{F}}_{\rm in(out)}(\omega)$ is the input (output) Langevin force vector. These vectors contain the input (output) operators associated to the different channels, i.e. the antenna (both the radiative and non-radiative components, $\aop_{\rm r, in (out)} (\omega)$ and $ \aop_{\rm nr, in(out)} (\omega)$, respectively) and the spin ensembles ($\bop_{\mu,{\rm in(out)}} (\omega)$). 

Since we focus on coherent reflection spectra where the signal enters and is sampled through the radiative port of the antenna, $S_{11}(\omega)$, we assume that $\expec{\hat a_{\rm nr, in}}= \expec{\hat b_{\mu,{\rm in}}}=0$. 
Therefore, taking the expectation value of \eqref{eq:meth-output_vs_input}, where $\expec{\hat{\mathbf{F}}_{\rm in}} = (\sqrt{\gamma_{\rm r}} \, \expec{\hat a_{\rm r, in}}, 0 \,,\, \dots)^T$, and applying the input-output relation for the radiative port, $\hat{a}_{\rm r, out}(\omega)= \hat{a}_{\rm r, in}(\omega)-\sqrt{\gamma_{\rm r}} \hat{a}(\omega)$, we derive the reflection coefficient
\begin{equation} \label{eq:meth-S11} 
\begin{aligned}
    S_{11}(\omega) & = \frac{\expec{\hat a_{\rm r, out}(\omega)}}{\expec{\hat a_{\rm r,in}(\omega)}} = \frac{\frac{-\gamma_{\rm r}+\gamma_{\rm nr}}{2}-i (\omega-\omega_0)+\sum^{N}_{\mu=1}\frac{g_\mu^2}{\frac{\gamma_{{\rm s} \mu}}{2}-i(\omega-\omega_{{\rm s}\mu})}}{\frac{\gamma_{\rm r}+\gamma_{\rm nr}}{2}-i (\omega-\omega_0)+\sum^{N}_{\mu=1}\frac{g_\mu^2}{\frac{\gamma_{{\rm s} \mu}}{2}-i(\omega-\omega_{{\rm s}\mu})}} \\
    & = 1- \frac{\gamma_{\rm r}}{\frac{\gamma_{\rm r}+\gamma_{\rm nr}}{2}-i (\omega-\omega_0)+\sum^{N}_{\mu=1}\frac{g_\mu^2}{\frac{\gamma_{{\rm s} \mu}}{2}-i(\omega-\omega_{{\rm s}\mu})}}\,.
\end{aligned}
\end{equation}
This equation serves as a model for fitting experimental reflectivity spectra. Notably, as pointed out in \secref{sec_theory} and in Refs.~\cite{SweeneyPhysRevA.102.063511,lamberto2024superradiantquantumphasetransition,Rao2024NaturePhysics}, the resonances of the reflection scattering coefficient $S_{11}(\omega)$ corresponds to the zeros $\Omega_j$ of the characteristic polynomial of the non-Hermitian Hamiltonian $\hat{H}_{\rm res} / \hbar = \hat{\boldsymbol{\alpha}}^\dagger (\mathbf{A} - i \mathbf{\Gamma}/2) \hat{\boldsymbol{\alpha}}$.
The numerator has the same structure as the denominator, differing only in the sign of the damping rate associated with the input-output channel, $\gamma_{\rm r}$. By introducing the effective decay matrix $\tilde{\bf \Gamma}$, which retains the structure of $\bf \Gamma$ but with the sign of $\gamma_{\rm r}$ reversed, the numerator can be interpreted as the characteristic polynomial of the effective non-Hermitian Hamiltonian $\hat{H}_{\rm RZ}/\hbar = \hat{\boldsymbol{\alpha}}^\dagger (\mathbf{A} - i \tilde{\mathbf{\Gamma}}/2) \hat{\boldsymbol{\alpha}}$, whose eigenfrequencies are $\tilde{\Omega}_j$.

\subsection{Derivation of the effective Hamiltonian in the strong coupling regime}

In this section we present the derivation for a single spin ensemble of the effective non-Hermitian Hamiltonian in the strong coupling regime.
Starting from the quantum Langevin equations for the photon and the collective spin operators presented in the Supplementary Information, we apply the unitary transformation $\bf U$ defined in \secref{sec_theory} to rotate in the polariton basis, i.e. $\hat{\mathbf{P}} = \mathbf{U} \, \hat{\boldsymbol{\alpha}}$.
Therefore, we obtain the following equations of motion for the polaritonic bosonic operators
\begin{equation} \label{eq:meth-EOM_P}
    \partial_t \hat{P}_j =  -i \bar{\omega}_j \hat{P}_j-\int_{-\infty}^{\infty}\,d\omega \, U_{j1} \sum_{\rm k={\rm r, nr}}\sqrt{\frac{\gamma_{\rm k}}{2\pi}}\, \hat{c}_{\rm k}(\omega) -\int_{-\infty}^{\infty}\,d\omega \, U_{j2}\, \sqrt{\frac{\gamma_{\rm s}}{2\pi}}\,\hat{d}(\omega) \, ,
\end{equation}
where $\hat{c}_k$ and $\hat{d}$ are the baths' bosonic operators, which are used to construct the corresponding input (output) operators, $\hat{a}_{\rm k,{\rm in(out)}}$ and $\hat{b}_{\rm in(out)}$ (see Supplementary Information). $\bar{\omega}_j$ and $U_{jk}$ are the polaritonic eigenfrequencies and Hopfield coefficients, respectively, as defined in the main text.

Following the approach outlined in Refs.~\cite{walls2008quantum,gardiner2004quantumpp}, we substitute the formal solutions of the equations of motion for $\hat{c}_k(\omega)$ and $\hat{d}$ into \eqref{eq:meth-EOM_P}. By taking the expectation value of the resulting expression, we obtain
\begin{align} \label{eq:meth-EOM_P_expect}
    \partial_t \, p_j(t) = &  -i \bar{\omega}_j \, p_j(t)- \frac{\gamma_j}{2}\, p_j(t) - U_{j1} U_{m1}^* \,\frac{\gamma_{\rm r} + \gamma_{\rm nr}}{2}\, p_m(t) \nonumber \\
    & - U_{j2} U_{m1}^* \,\frac{\gamma_s}{2}\,p_m(t)+\sqrt{\gamma_{\rm r}}\, U_{j1}\, a_{\rm in}(t) \, ,
\end{align}
where $m$ is the complementary index of $j$, i.e. $m=2$ when $j=1$ and viceversa. In addition, we introduced the definitions $p_j (t) = \langle \hat{P}_j (t) \rangle$, $\gamma_j = (\gamma_{\rm r} + \gamma_{\rm nr}) \abs{U_{j1}}^2 + \gamma_s \abs{U_{j2}}^2$ and $a_{\rm in}(t) = \expec{\hat{a}_{\rm r, in} (t)}$, which represents the coherent input signal. 
Specifically, in the derivation of \eqref{eq:meth-EOM_P_expect}, we took into account that the system has a coherent feeding only through the antenna radiative channel, thus implying $\expec{\hat{a}_{\rm nr, in}} = \langle \hat{b}_{\rm in} \rangle = 0$.
\eqref{eq:meth-EOM_P_expect} can be rewritten in the frequency domain as
\begin{align} \label{eq:meth-EOM_P_expect_freq}
    -i \omega \, p_j(\omega) = &  -i \bar{\omega}_j \, p_j(\omega)- \frac{\gamma_j}{2}\, p_j(\omega) - U_{j1} U_{m1}^* \,\frac{\gamma_{\rm r} + \gamma_{\rm nr}}{2}\, p_m(\omega) \nonumber \\
    & - U_{j2} U_{m1}^* \,\frac{\gamma_{\rm s}}{2}\,p_m(\omega)+\sqrt{\gamma_{\rm r}}\, U_{j1}\, a_{\rm in}(\omega) \, .
\end{align}

By imposing the perfect absorption condition, $a_{\rm out}(\omega)= 0$, we obtain from the input-output relations the explicit expression $a_{\rm in}(\omega) =  \sqrt{\gamma_{\rm r}}\, ( U_{11}\, p_1 (\omega)+ U_{12}\, p_2 (\omega))$, which can be inserted in \eqref{eq:meth-EOM_P_expect_freq}. Furthermore, for $\omega \approx \bar{\omega}_j$, only $p_j$ is significantly excited by the input field, while the terms involving the other polariton mode can be safely neglected in the strong coupling regime, due to the great separation of the spectral lines. Hence, the effective dynamics in the time domain can be written as
\begin{equation}
    \partial_t \, p_j(t) = - i \bar{\omega}_j \, p_j(t) - \frac{\bar{\gamma}_j}{2}\,p_j(t)\, ,
\end{equation}
where $\bar{\gamma}_j = (-\gamma_{\rm r}+\gamma_{\rm nr}) \, \abs{U_{j1}}^2 + \gamma_{\rm s} \, \abs{U_{j2}}^2$, as in \eqref{eq:gamma_bar_strong}. 
These equations of motions can be derived by the effective non-Hermitian Hamiltonian
\begin{equation}
    \hat{H}_{\rm RZ} = \sum_{j=1,2} \left( \bar{\omega}_j - i \frac{\bar{\gamma}_j}{2} \right) \hat{P}_j^\dagger \hat{P}_j \, ,
\end{equation}
which coincides with \eqref{eq:H_eff_polariton}. This procedure can be easily generalized in the case of multiple spin resonances, as in the strong coupling regime each photon-spin anti-crossing is separated from the others.

\subsection{Experimental Set up}\label{subsec_setup}

We use a planar superconducting lumped-element LC microwave resonator made of superconducting Niobium films (thickness, 50 nm) on sapphire substrate (thickness, 420 $\mu$m) as shown in \figref{Fig1}. The resonator has a small inductive loop to enhance the generated microwave magnetic field, which is coupled to a large interdigitated capacitance. The resonator works in reflection mode, at a fundamental frequency $\omega_{0}/2\pi\,\approx\,9.9\,$GHz, and it is coupled to the input-output line through an antenna, whose position can be adjusted at room temperature, before cooling the sample. The chip carrying the resonator is loaded into a cylindrical copper waveguide sample holder hosting the antenna in his bottom \cite{riegerPHYSREVAPPL2023,riegerNATMAT2023}. Preliminary characterization of the empty resonator are reported in Supplementary Information.

All experiments are carried out inside a Qinu Sionludi dilution refrigerator (base temperature 20 mK) equipped with a three-axial superconducting magnet and microwave lines and electronics \cite{riegerPHYSREVAPPL2023,riegerNATMAT2023}. The input signal is attenuated by 60 dB inside the cryostat before reaching the sample box with the antenna, while the output line hosts a cryogenic High Electron Mobility Transition (HEMT) amplifier (Low Noise Factory, 37 dB gain). The signal is further amplified at room temperature before acquisition. The complex reflection scattering parameter, $S_{11}$, is measured with a Vector Network Analyzer (VNA) for different values of the static magnetic field applied, obtaining the 2D maps shown in \figref{fig:single_spin} and \figref{fig:multi_spin}. The input power at the position of the antenna is between -130 and -120 dBm, corresponding to average single microwave photon into the resonator (see Supplementary Information). The cylindrical box and the resonator are aligned into the magnet in order to have the static magnetic field along the plane of the chip, \textit{i.e.} in a in-plane configuration (\figref{Fig1} (a)). All the measured reflection scattering parameters have been normalized over the average value of the signal baseline measured off-resonance.

\subsection{Molecular Spin Samples}\label{subsec_molspin}

We use two samples of the molecular compounds shown in \figref{Fig1} (c). 
The first one is a diluted solid dispersion of $\alpha,\gamma$-bisdiphenylene-$\beta$-phenylallyl (BDPA, for short) organic radical into a Polystirene matrix, with a spin density of $\approx 1\cdot 10^{15}$ spin/mm$^3$. The sample was prepared as described in \cite{bonizzoniAPPLMAGNRES2023} and then cut into a rectangular shape with size $\approx 1.5 \times 1\,\text{mm}^2$. The sample is placed on the resonator as shown in \figref{Fig1} (a). Each molecule has electronic spin $S=1/2$, which is due to its single unpaired electron \cite{lenzADVMAT2021,bonizzoniAPPLMAGNRES2023}. Pure BDPA samples typically have antiferromagnetic exchange interaction occuring below 10 K \cite{lenzADVMAT2021,duffyJOURNCHEMPHYS1972,burgessJOURNAPPLPHYS1962}, with a Curie-Weiss temperature between $T_{C}=-8\,$K and $T_{C}=-1\,$K, strongly dependent by the spin concentration and by the solvent or matrix used \cite{lenzADVMAT2021,duffyJOURNCHEMPHYS1972,burgessJOURNAPPLPHYS1962}. Although there are no reports for BDPA diluted in Polystyrene, due to the relatively high concentration, we can expect that a residual antiferromagnetic interaction still occurs among molecules. Therefore, our BDPA sample constitutes a prototypical TLS collection with essentially negligible magnetic anisotropy and no hyperfine splitting, which gives a single transition frequency \cite{lenzADVMAT2021,bonizzoniAPPLMAGNRES2023}. 

The other sample is a single crystal of VOTPP with 2\% concentration in its isostructural diamagnetic analog, TiO(TPP). Each molecule has an electronic spin $S\,=\,1/2$ ground state and an additional hyperfine splitting due to the interaction with the $I=7/2$ nuclear spin of the $^{51}$V ion (natural abundance: 99.75\%). This results in a multiplet with eight $\{S_{z},I_{z} \}$ electronuclear transitions, which can be exploited as eight independent spin ensembles. The magnetic properties and the electron spin resonance spectroscopy of this molecule have been previously reported in \cite{yamabayashiJACS2018}. In particular, the hyperfine tensor shows uniaxial anisotropy with parallel component $A_{\parallel}=477\,MHz = 23\,\text{mK}$ and perpendicular component $A_{\perp}=168\,MHz = 8\,\text{mK}$ \cite{yamabayashiJACS2018}, respectively. These relatively-large values, combined with the temperature of the experiments and the frequency of the resonator, give different thermal population on the eight lines, thus decreasing the coupling rate $g_{\mu}$ for increasing line number $i$ (see Supplementary Information). The position of the VOTPP crystal (from \#A to \#D in \figref{Fig1} (a)) is adjusted before each cooldown. Due to the experimental configuration used and the orientation of the molecules inside the unit cell, all molecules experience the same static magnetic field, which lies on the TPP plane and it is nearly perpendicular to the direction of the $V=0$ double bound.

\backmatter

\bmhead{Supplementary information}

Supplementary Information contains extended theoretical derivations, additional experimental results as well as all fit parameters.

\bmhead{Acknowledgements}

We thank Dr. Johan van Tol (National High Magnetic Field Laboratory, Florida, USA) for the preparation of BDPA sample.\\
We thank Dr. Filippo Troiani (CNR Istituto Nanoscienze, Centro S3, Modena, Italy) for stimulating discussion. 
C.B. acknowledges the International Excellence Grants Program of KIT for support thorugh the University of Excellence concept.

\section*{Declarations}


\begin{itemize}
\item Funding: This work was funded by the National Quantum Science and Technology Institut (NQSTI) PE00000023 SPOKE 5- call N. 1 Project "Addressing molecular and Spins with MIcrowave puLsEs through Superconducting circuits for QUantum Information Processing (SMILE-SQUIP)", by the US Office of Naval Research award N62909-23-1-2079 project "Molecular Spin Quantum Technologies and Quantum Algorithms", and by the Army Research Office
(ARO) (Grant No. W911NF-19-1-0065).
This work was also supported by the International Excellence Grants Program of KIT with funds granted to the University of Excellence concept.

\item Conflict of interest/Competing interests: All authors declare no competing interests. 
\item Ethics approval and consent to participate:  n.a.
\item Consent for publication: n.a.
\item Data availability: Data are available from corresponding author upon reasonable request.
\item Materials availability: n.a.
\item Code availability:  n.a.
\item Author contribution: The experiment was conceived and designed by C.B., A.G., M.A. The theoretical framework has been developed by D.L., S.N. and S.S. The measurement were carried out by C.B. and S.G. at Karlsruhe Institute of Technology, unbder the supervision of W.W.. The resonator were designed by D.R.. Data fitting and simulation were done by D.L. and S.N. (with equal contributions). The VOTPP sample was prepared by F.S., who carried out also the EasySpin simulation of VOTPP. The manuscript was written by C.B. and D.L. with inputs from all coauthors. All authors contributed to the discussion of the results. The manuscript has been revised by all authors before submission.  

\end{itemize}

\noindent

\bigskip
\begin{flushleft}%
Editorial Policies for:

\bigskip\noindent
Springer journals and proceedings: \url{https://www.springer.com/gp/editorial-policies}

\bigskip\noindent
Nature Portfolio journals: \url{https://www.nature.com/nature-research/editorial-policies}

\bigskip\noindent
\textit{Scientific Reports}: \url{https://www.nature.com/srep/journal-policies/editorial-policies}

\bigskip\noindent
BMC journals: \url{https://www.biomedcentral.com/getpublished/editorial-policies}
\end{flushleft}

\bibliography{biblio.bib}



\section*{Supplementary material}

SUPPLEMENTARY INFORMATION OF:

Observation of Perfect Absorption in Hyperfine Levels of Molecular Spins with Hermitian Subspaces

Claudio Bonizzoni[1,2*], Daniele Lamberto[3], Samuel Napoli[3], Simon Günzler[4], Dennis Rieger[4], Fabio Santanni[5], Alberto Ghirri[2], Wolfgang Wernsdorfer[4], Salvatore Savasta[3], Marco Affronte[1,2]

[1*]Dipartimento di Scienze Fisiche, Informatiche e Matematiche, Università di Modena e Reggio Emilia, via G. Campi 213/a, Modena, 41125, Italy.
[2]Istituto Nanoscienze, CNR, via G. Campi 213/a, Modena, 41125, Italy.
[3]Dipartimento di Scienze matematiche e informatiche, scienze fisiche e scienze della terra, Università degli Studi di Messina, Via Salita Sperone, c.da Papardo, Messina, 98166, Italy.
[4]Physikalisches Institut, Karlsruhe Institute of Technology, Wolfgang-Gaede-Str. 1, Karlsruhe, D-76131, Germany.
[5]Dipartimento di Chimica Ugo Schiff, Università degli Studi di Firenze, via della Lastruccia 3, Sesto Fiorentino (FI), 50019, Italy.

*Corresponding author(s). E-mail(s): claudio.bonizzoni@unimore.it;

# Contents

# 1 Quantum Langevin Equations for the multiple spin ensemble coupled to the electromagnetic resonator

Here, we derive the quantum Langevin equations leading to the reflection scattering parameter in Eq. (3) of the main text. The total Hamiltonian $\hat{H}_T$ for either the VOTPP or the BDPA sample, including their interaction with the electromagnetic resonators and all the reservoir channels, is given by:

$$\hat{H}_T = \hat{H}_S + \hat{H}_B + \hat{H}_I. \tag{S.1}$$

The system Hamiltonian $\hat{H}_S$ is provided in Eq. (1) of the main text. The bare reservoir Hamiltonian is given by

$$\hat{H}_B = \hbar \int_{-\infty}^{+\infty} d\omega\, \omega \left( \sum_{\text{k=r,nr}} \hat{c}_{\text{k}}^\dagger(\omega)\hat{c}_{\text{k}}(\omega) + \sum_{\mu=1}^{N} \hat{d}_\mu^\dagger(\omega)\hat{d}_\mu(\omega) \right), \tag{S.2}$$

where $\hat{c}_{\text{nr}}$ and $\hat{c}_{\text{r}}$ are the bosonic annihilation operators associated with the internal and radiative losses of the resonator, respectively. Note that $\hat{d}_\mu$ represents the annihilation operator describing the reservoir field of the $\mu$-th spin ensemble. Finally, the interaction of the system and the resonator with their corresponding reservoirs is described by (under the rotating wave approximation (RWA) [1, 2])

$$\hat{H}_I = i\hbar \int_{-\infty}^{+\infty} d\omega \left[ \sum_{\text{k=r,nr}} \sqrt{\frac{\gamma_{\text{k}}}{2\pi}} \left( \hat{c}_{\text{k}}(\omega)\hat{a}^\dagger - \hat{c}_{\text{k}}^\dagger(\omega)\hat{a} \right) + \sum_{\mu=1}^{N} \sqrt{\frac{\gamma_{\text{s}\mu}}{2\pi}} \left( \hat{d}_\mu(\omega)\hat{b}_\mu^\dagger - \hat{d}_\mu^\dagger(\omega)\hat{b}_\mu \right) \right], \tag{S.3}$$

where the decay rates are defined in the main text. Following a standard approach to open quantum systems [1, 2], we derive a set of quantum Langevin equations by substituting the solutions of the reservoir operators into the Heisenberg equations of motion for the system operators (the spin ensemble and the resonator):

$$\frac{d}{dt}\hat{\boldsymbol{\alpha}}(t) = -i\mathbf{A}\hat{\boldsymbol{\alpha}}(t) - \frac{1}{2}\boldsymbol{\Gamma}\hat{\boldsymbol{\alpha}}(t) + \hat{\mathbf{F}}_{\text{in}}(t), \tag{S.4}$$

$$\frac{d}{dt}\hat{\boldsymbol{\alpha}}(t) = -i\mathbf{A}\hat{\boldsymbol{\alpha}}(t) + \frac{1}{2}\boldsymbol{\Gamma}\hat{\boldsymbol{\alpha}}(t) + \hat{\mathbf{F}}_{\text{out}}(t). \tag{S.5}$$

In the last expressions, we introduced the block matrices

$$\hat{\boldsymbol{\alpha}} = \begin{pmatrix} \hat{a}(t) \\ \hat{b}_1(t) \\ \vdots \\ \hat{b}_N(t) \end{pmatrix}, \quad \mathbf{A} = \begin{pmatrix} \omega_0 & \mathbf{g}^T \\ \mathbf{g} & \boldsymbol{\omega_s} \end{pmatrix}, \quad \boldsymbol{\Gamma} = \begin{pmatrix} \gamma_{\text{r}} + \gamma_{\text{nr}} & 0 \\ 0 & \boldsymbol{\gamma_s} \end{pmatrix}, \tag{S.6}$$

where $\boldsymbol{\omega_s}$ and $\boldsymbol{\gamma_s}$ are diagonal block matrices with the $\mu$-th elements given by $\omega_{s\mu}$ and $\gamma_{s\mu}$ (where $\mu$ ranges from 1 to $N$), respectively. Additionally, we defined

$$\boldsymbol{g}^T = (g_1, ..., g_N) \,, \tag{S.7}$$

and the input and output field vectors as

$$\hat{\mathbf{F}}_{\text{in}}(t) = \begin{pmatrix} \sqrt{\gamma_r}\hat{a}_{r,\text{in}}(t) + \sqrt{\gamma_{nr}}\hat{a}_{nr,\text{in}}(t) \\ \sqrt{\gamma_{s\mu}}\hat{b}_{1,\text{in}}(t) \\ \vdots \\ \sqrt{\gamma_{sN}}\hat{b}_{N,\text{in}}(t) \end{pmatrix} \,,\; \hat{\mathbf{F}}_{\text{out}}(t) = \begin{pmatrix} \sqrt{\gamma_r}\hat{a}_{r,\text{out}}(t) + \sqrt{\gamma_{nr}}\hat{a}_{nr,\text{out}}(t) \\ \sqrt{\gamma_{s\mu}}\hat{b}_{1,\text{out}}(t) \\ \vdots \\ \sqrt{\gamma_{sN}}\hat{b}_{N,\text{out}}(t) \end{pmatrix} \,. \tag{S.8}$$

where the components of the input vector are given by

$$\begin{aligned} \hat{a}_{r,\text{in}}(t) &= \frac{1}{\sqrt{2\pi}} \int_{-\infty}^{+\infty} d\omega \, e^{-i\omega(t-t_0)} \hat{c}_r(\omega; t_0) \,, \\ \hat{a}_{nr,\text{in}}(t) &= \frac{1}{\sqrt{2\pi}} \int_{-\infty}^{+\infty} d\omega \, e^{-i\omega(t-t_0)} \hat{c}_{nr}(\omega; t_0) \,, \\ \hat{b}_{\mu,\text{in}}(t) &= \frac{1}{\sqrt{2\pi}} \int_{-\infty}^{+\infty} d\omega \, e^{-i\omega(t-t_0)} \hat{d}_\mu(\omega; t_0) \,. \end{aligned} \tag{S.9}$$

where $t_0 < t$ is the initial time (input) chosen to solve the equations of motion for the reservoir field operators. Similarly, the output field vector entries have the same form as in Eq. (S.9), except that $t_0$ is replaced by $t_f$ ($t < t_f$, output). Rewriting the Langevin equations Eq. (S.4) and Eq. (S.5) in the frequency domain allows us to express $\mathbf{F}_{\text{out}}(\omega)$ in terms of $\mathbf{F}_{\text{in}}(\omega)$:

$$\hat{\mathbf{F}}_{\text{out}}(\omega) = \left[-\frac{1}{2}\boldsymbol{\Gamma} - i(\omega \mathbf{I} - \mathbf{A})\right]\left[\frac{1}{2}\boldsymbol{\Gamma} - i(\omega \mathbf{I} - \mathbf{A})\right]^{-1} \hat{\mathbf{F}}_{\text{in}}(\omega) \,, \tag{S.10}$$

where $\mathbf{I}$ is the identity matrix. Moreover, the input-output relationships are derived by subtracting Eq. (S.4) to Eq. (S.5) in the frequency domain, namely

$$\hat{\mathbf{F}}_{\text{out}}(\omega) = -\boldsymbol{\Gamma}\hat{\boldsymbol{\alpha}}(\omega) + \hat{\mathbf{F}}_{\text{in}}(\omega) \,. \tag{S.11}$$

From this equation follows the input-output relation for the radiative port, $\hat{a}_{r,\text{out}}(\omega) = \hat{a}_{r,\text{in}}(\omega) - \sqrt{\gamma_r}\,\hat{a}(\omega)$, which is used to calculate the complex scattering coefficient (see Methods).

# 2 Additional Experimental Details

## 2.1 Resonators and Cryostat

Each chip used in the experiments is made of sapphire with dimensions $3 \times 10 \, \text{mm}^2$ and thickness $420 \, \mu$m. The short side of the chip carries two planar microwave resonators, namely Res #1 and Res #2, working in reflection mode and coupled to the same feeding antenna, whose position can be adjusted at room temperature, as in Fig. S1 and Fig. (1) of main text. The two resonators are made of superconducting Niobium films (thickness, 50 nm). The geometry is a lumped-element LC resonator, in which a the interdigitated part acts as a large capacitance and the central loops give a small inductance, giving large microwave currents around the loop position. The width of the superconducting niobium strip is $w = 10 \, \mu m$, while the interspace between each part of the capacitors is $w' = 20 \, \mu m$. The length of the resonator is $l_1 = 1$ mm, while its distance from the edge of the chip is $d_1 = 190 \, \mu$m. Res #1 and Res #2 are designed to have two fundamental resonant frequency $\omega_{0,1}/2\pi = \nu_{0,1} \approx 9.9 \, \text{GHz}$ and $\omega_{0,2}/2\pi = \nu_{0,2} \approx 11 \, \text{GHz}$, respectively. The large detuning between their bare resonance frequencies makes Res #1 and #2 fully independent each other. In this work only data obtained from Res #1 are presented.

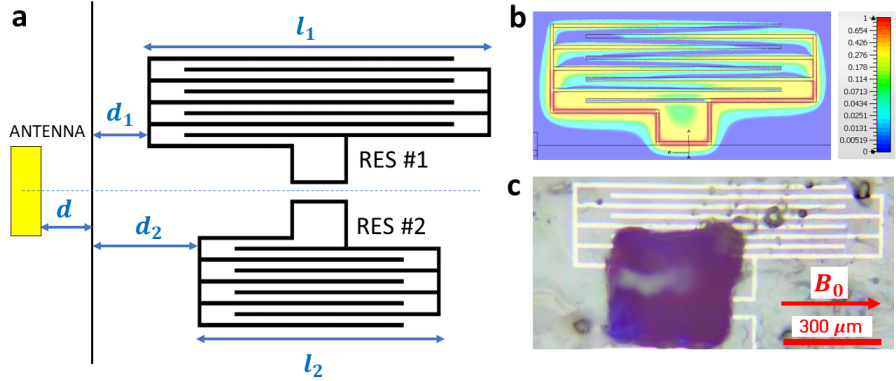

**Figure S1**: **a** Sketch of the two resonators and of their position with respect to the input-output antenna. **b** Electromagnetic simulation of the bare resonator showing the distribution of the magnitude of the magnetic component of the microwave field. The colorscale on right is normalized on the maximum field value, with red corresponding to maximum and blue to 0. **c** Photo of the planar niobium lumped element resonator with a VOTPP crystalline sample on it. The sample has been partially moved to show the underlying inductive loop of the resonator. Red arrow shows the direction of the applied static magnetic field, $B_0$.

The chip is loaded into the same sample holder described in Refs. [3–5], which is essentially a cylindrical copper waveguide hosting the antenna in its bottom part. The

sample holder is cooled inside a Qinu Sionludi dilution refrigerator with a base temperature of 20 mK, and equipped with three-axial superconducting coils and microwave lines and electronics [3, 4]. The chip and the sample holder are oriented the way that the main coil generates the static magnetic field in the plane of the resonators.

## 2.2 Estimation of the microwave photon number

We estimate the average number of microwave photons, $n_{av}$, in the resonator as reported in [6], using Eq. (S.12):

$$n_{av} = \frac{P_{in} Q_{L,1} 10^{-\frac{IL_1}{20}}}{\pi h \nu_{0,1}^2}. \tag{S.12}$$

Here, $P_{in}$ is the input power at the input-output antenna, $Q_{L,1}$ is the loaded quality factor, $IL_1$ is the insertion loss of the resonator Res. #1 and $\nu_{0,1}$ its resonant frequency. $h$ is Planck's constant. The calculation as a function of the input power is shown in Fig. S2. The input power values used in this work give an average unitary photon number into the resonator.

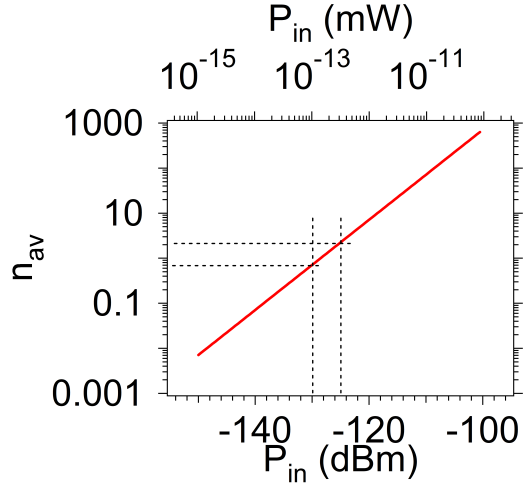

**Figure S2**: Average number of microwave photons in Res. #1 calculated with Eq. (S.12) as a function of the input power at the antenna. Vertical dashed lines show the range of power values used in this work, corresponding to a single photon on average filling the resonator.

Due to the low photon number used, the temperature and the resonant frequency we estimate the contribution of thermal photons to the photon number. To this end, we consider the resonator as a black body at a given finite temperature, $T$, and we consider Planck's law:

5

$$P_{th} = \frac{8\pi h\nu^3}{c^3} \frac{1}{e^{\frac{h\nu}{\kappa_B T}} - 1}. \tag{S.13}$$

Eq. (S.13) gives the power emitted per unit of frequency, $\nu$, unit of solid angle and unit of surface. Here, $h$ is again Planck's constant, while $\kappa_B$ is Boltzmann's constant and $c$ is the speed of light in vacuum. The number of thermal photons per unit of time, frequency, solid angle and surface, $n_{th}$ (which corresponds to the number of photons per unit of solid angle and surface), is obtained by dividing Eq. (S.13) by the energy of a single photon, $h\nu$. This brings to:

$$n_{th} = \frac{P_{th}}{h\nu} = \frac{8\pi\nu^2}{c^3} \frac{1}{e^{\frac{h\nu}{\kappa_B T}} - 1}. \tag{S.14}$$

Fig. S3 shows the results obtained with Eq. (S.14) as a function of frequency for different experimental temperature values. The results show that at the resonant frequency of Res. #1 ($\approx 9.9$ GHz, vertical dashed line) the power associated to thermal photon emission is much lower than the input microwave one. Moreover, the corresponding number of photons per unit of solid angle and surface is negligible with respect to the average microwave photon number of the microwave tone. Therefore, the effect of thermal photons can be neglected in our measurements.

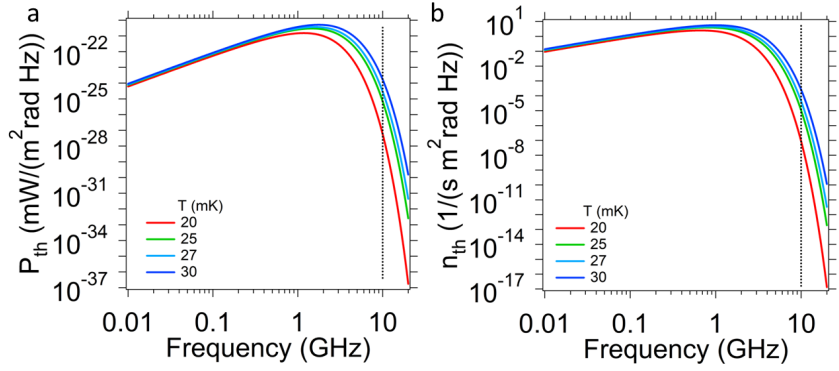

**Figure S3**: **a** Power per unit of frequency, solid angle and surface as a function of frequency calculated with Eq. (S.13) for different experimental temperatures values used. **b** Number of photons per unit of solid angle and surface calculated with Eq. (S.14) as a function of frequency. In both a and b vertical dashed lines indicated the frequency of Res. #1.

# 3 Additional Data

## 3.1 Data for empty resonators

An example of complex reflection scattering parameter (amplitude, $|S_{11}|$, and phase) measured at 30 mK on Resonator Res #1, at microwave power equivalent to single photon level and in zero magnetic field is shown in Fig. S4. Black line is a fit performed by means of Eq. (3) of main paper keeping $g_\mu = 0$ for any $\mu$.

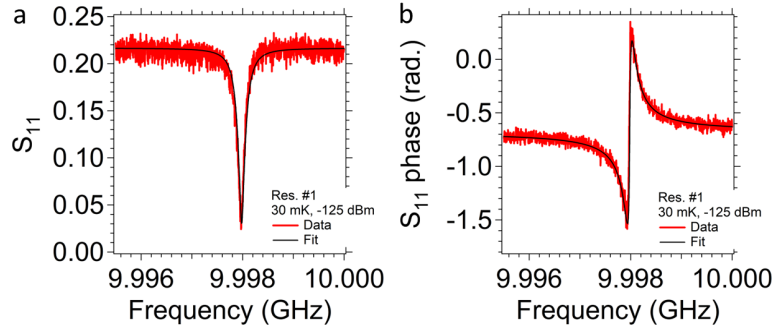

**Figure S4**: Reflection and phase signals measured at 30 mK and single photon powers for a resonator as the one shown in Fig. 1 of main text. Dashed line are fits based on Eq. (3) of main paper with fixed $g_\mu = 0$ for any $\mu$.

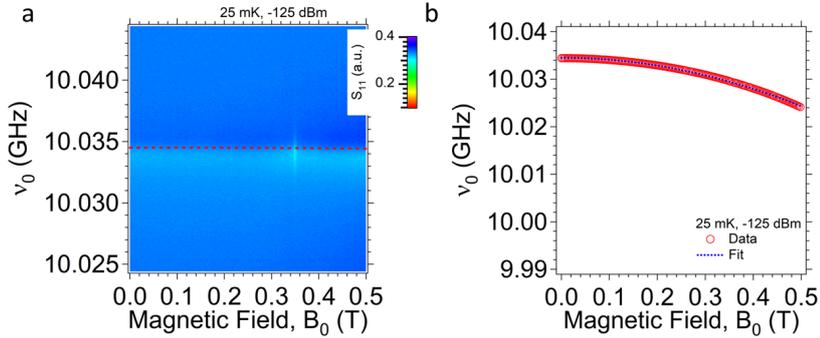

**Figure S5**: 2D reflection map **a** and resonant frequency as a function of the static in-plane magnetic field **b** for resonator Res. #1. Dashed line is a fit based on Eq. (S.15).

Fig. S5 shows the reflection map for the empty resonator Res. #1 taken at 25 mK for different applied in-plane magnetic field values. The resonance can be measured at least up to $B_0 = 0.5$ T. No signal, except for a very weak one due to magnetic

7

impurities, and no dips with zero reflection are visible in the range in which the resonance with spins is expected ($\approx 0.34$ T). The resonant frequency decreases as a function of the static magnetic field as expected from the magnetic field response of niobium under applied in-plane field. While this change in frequency is almost negligible for the single-ensemble results it becomes more heavier on the multiple ensemble data due to the larger field span used. We take into account this this effect by including in our model the dependence of the resonant frequency on the static magnetic field correction of the resonant frequency of the resonator according to:

$$\omega_0(B_0)/2\pi = \nu_0(B_0) = \bar{\nu}_0 - aB_0^2. \tag{S.15}$$

Here, $\bar{\nu}_0$ is the bare resonator frequency at zero field and $a$ a phenomenological parameter. The value of $a$ is obtained by fitting the in-plane dependence of the frequency of the resonator for an empty resonator (Fig. S5).

### 3.2 Tuning antenna position on BDPA

Figure S6 shows 2D normalized reflection maps measured for the BDPA sample for different positions of the input-output antenna. The BDPA sample is placed in similar positions over different experiments to obtain strong coupling regimes similar to the one of Fig. 2 of main paper. The position of the antenna is changed at room temperature before cooldown. We investigate the effect of four different positions, named as Ant. #1 to Ant. #4 in Fig. S6, and corresponding to different decreasing coupling strengths between the antenna and the resonator. An avoided crossing is always visible in the 2D map around resonance, as in Fig. 2 of the main paper. The changes in the bare resonator frequency can be attributed to the different cooldown cycles and to the different coupling with the input-output antenna. Specifically, in Fig. S6 (a, d), $|S_{11}|$ approaches zero for each polariton, indicating the occurrence of perfect absorption away from the central resonance. This is confirmed by the occurence of two zeroes in the corresponding imaginary part of the eigenfrequencies, $\tilde{\Omega}_j$ (Fig. S6 (c, f)). In contrast, antenna positions Ant. #3 and Ant. #4 (Fig. S6 (g, l)) exhibit an anticrossing with higher reflection values and no signature of perfect absorption. The corresponding fit results confirm a significant reduction in $\gamma_{\rm r}$ relative to Fig. S6 (a, d) (see tables in Sec. 4). This reduction influences the imaginary parts of the eigenfrequencies $\tilde{\Omega}_j$, giving no zeroes around resonance.

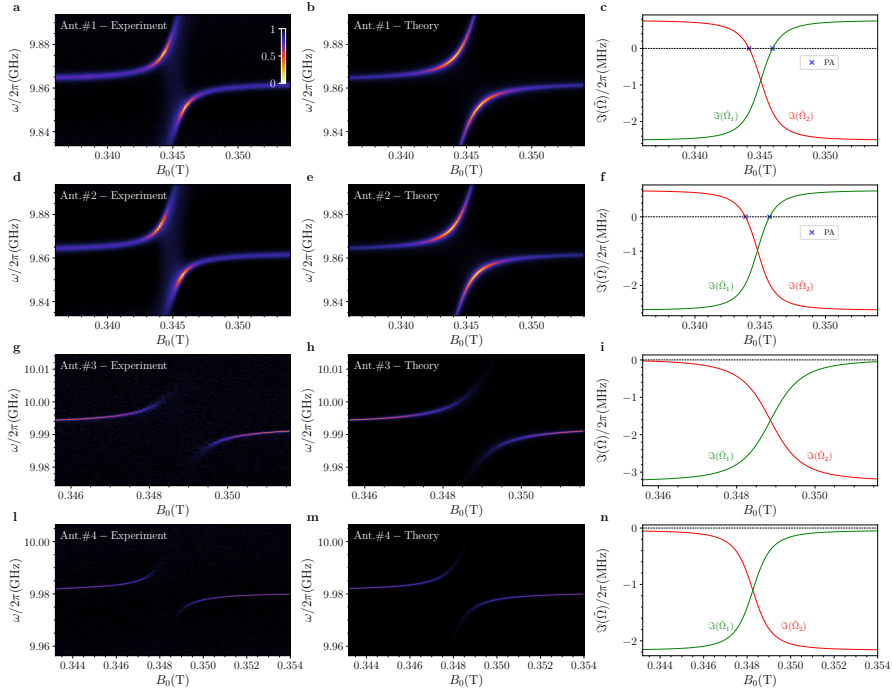

**Figure S6**: **a, d, g, l** Normalized reflection maps ($|S_{11}|$) measured on BDPA for various positions of the input-output antenna, labeled Ant. #1 to Ant. #4. These correspond to decreasing coupling strengths between the antenna and the resonator (i.e., decreasing $\gamma_r$). Note that the map for Ant. #1 is identical to the one shown in Fig. 2 (a) of main text. **b, e, h, m** Reflection maps obtained by fitting the data in **a, d, g, l** using Eq. 8 in Methods. **c, f, i, n** Imaginary parts of the eigenfrequencies $\tilde{\Omega}_j$, extracted from the fitted parameters, confirm the presence of perfect absorption in the cases shown in **a, d**, and its absence in **g, i**.

### 3.3 Additional Data for VOTPP

Fig. S7 shows the extended data set for the VOTPP crystal. By changing the position of the crystal on the resonator (see Fig. 1 of main text) it is possible to tune the coupling between spins and resonator and to investigate different coupling regimes. Position #A (Fig. S7 (a,b,c)) and #D (Fig. S7 (l,m,n)) correspond to the results reported in main text, while positions #B and #C show intermediate sample positions. The overall coupling regime is mainly determined by the sample position, while a finer tuning is achieved through the different thermal populations of the lines. Our model perfectly fits all experimental data. The real, $\Re(\tilde{\Omega}_j)$, and imaginary, $\Im(\tilde{\Omega}_j)$, part of the polaritonic eingenstates correctly predict the positions and number of dips for all experimental dataset.

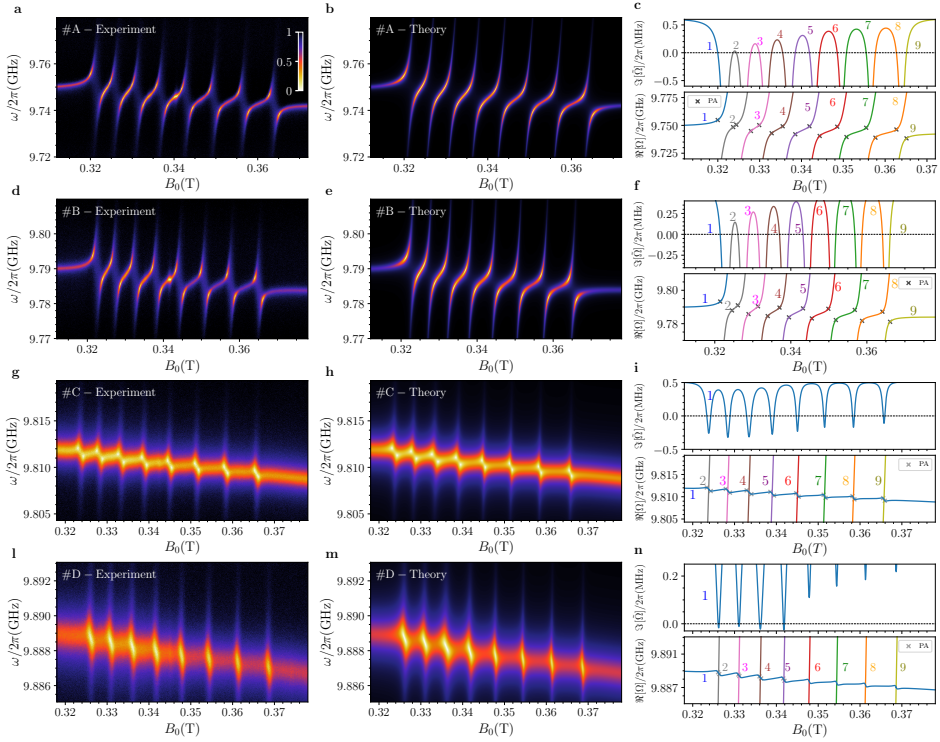

**Figure S7**: Extended dataset showing the results for all four positions of the VOTPP sample on the resonator (from #A to #D, see Fig. (1) of main text). For each position the measured 2D reflection map, the corresponding fit obtained with Eq. (3) of the main paper are shown. The third column show the Imaginary, $\Im(\tilde{\Omega}_j)$, and Real, $\Re(\tilde{\Omega}_j)$, part of the polaritonic frequencies. All dips are correctly predicted at the energies giving zero imaginary part.

## 4 Fit Parameters for all data sets

The following tables summarize the fit parameters extracted by means of Eq. (3) of main paper for all experimental datasets shown in the main text and in Sec. 3.3. Each table gives the parameters of the resonator obtained far from resonance and, then, the fit parameters for each $\mu^{th}$ spin ensemble.

### 4.1 Parameters for BDPA

Table S1, Table S2, Table S3 and Table S4 shows the fit parameters for the experimental dataset of BDPA given in Sec. 3.2. The parameters in Table S1 are the ones corresponding also to the data shown in Fig. (2) of main text. The coupling strength is larger than the corresponding linewidth further supporting that BDPA is in the strong coupling regime.

10

**Table S1:** $S_{11}(\omega)$ fitting parameters for the BDPA dataset shown in Fig. 2 of main text and in Fig. S6 (Ant. #1).

| Resonator parameters | | | |
|---|---|---|---|
| $a$(MHz/T$^2$) | $\bar{\nu}_0$(GHz) | $\gamma_r/2\pi$(MHz) | $\gamma_{nr}/2\pi$(MHz) |
| 0 | 9.863 | 1.67 | 0.13 |

| Spin ensemble parameters | | Coupling strength |
|---|---|---|
| $g_L$ | $\gamma_s/2\pi$(MHz) | $g/2\pi$(MHz) |
| 2.042 | 5.04 | 20.67 |

**Table S2:** $S_{11}(\omega)$ fitting parameters for the BDPA dataset shown in Fig. S6 (Ant. #2).

| Resonator parameters | | | |
|---|---|---|---|
| $a$(MHz/T$^2$) | $\bar{\nu}_0$(GHz) | $\gamma_r/2\pi$(MHz) | $\gamma_{nr}/2\pi$(MHz) |
| 0 | 9.863 | 1.64 | 0.094 |

| Spin ensemble parameters | | Coupling strength |
|---|---|---|
| $g_L$ | $\gamma_s/2\pi$(MHz) | $g/2\pi$(MHz) |
| 2.044 | 5.46 | 19.73 |

**Table S3:** $S_{11}(\omega)$ fitting parameters for the BDPA dataset shown in Fig. S6 (Ant. #3).

| Resonator parameters | | | |
|---|---|---|---|
| $a$(MHz/T$^2$) | $\bar{\nu}_0$(GHz) | $\gamma_r/2\pi$(MHz) | $\gamma_{nr}/2\pi$(MHz) |
| 0 | 9.993 | 0.16 | 0.056 |

| Spin ensemble parameters | | Coupling strength |
|---|---|---|
| $g_L$ | $\gamma_s/2\pi$(MHz) | $g/2\pi$(MHz) |
| 2.047 | 6.51 | 12.02 |

**Table S4:** $S_{11}(\omega)$ fitting parameters for the BDPA dataset shown in Fig. S6 (Ant. #4).

| Resonator parameters | | | |
|---|---|---|---|
| $a$(MHz/T$^2$) | $\bar{\nu}_0$(GHz) | $\gamma_r/2\pi$(MHz) | $\gamma_{nr}/2\pi$(MHz) |
| 0 | 9.981 | 0.06 | 0.132 |

| Spin ensemble parameters | | Coupling strength |
|---|---|---|
| $g_L$ | $\gamma_s/2\pi$(MHz) | $g/2\pi$(MHz) |
| 2.048 | 4.34 | 12.78 |

## 4.2 Parameters for VOTPP

Table S5, Table S6, Table S7, Table S8 show the fit parameters extracted from positions #A, #B, #C and #D, respectively (Fig. 4 of main text and Fig. S7). Overall, the coupling strength $g_\mu$ of each table decrease from position #A to #D, as a result of the lower amplitude of the microwave field of the resonator felt by the spins. Interestingly, for each sample position dataset, the fitted coupling strength decrease with increasing ensemble (line) index $\mu$, as a result of the different thermal population of each line, as discussed in main text. For position #A all fitted coupling strengths, $g_\mu$, are comparable to their corresponding linewidths, suggesting that the high cooperativity spin-photon coupling regime is achieved. A similar condition is observed also for position #B. Conversely, for position #C and #D each linewidth is significantly larger than its corresponding coupling strength, further suggesting that the weak spin-photon coupling regime is achieved. We attribute this overall increase of the spins decay-rates to the Purcell effect [7, 8]. Specifically, increasing such overlap converts a fraction of the spins decay-rates (spontaneous emission) into an increase of the emission in the resonator mode, which is described by the resonator-spins coupling rates.

**Table S5:** $S_{11}(\omega)$ fitting parameters for VOTPP in position #A, corresponding to the dataset shown in Fig. 4 of main text and in Fig. S7 (a,b,c).

| Resonator parameters | | | |
|---|---|---|---|
| $a$(MHz/T$^2$) | $\bar{\nu}_0$(GHz) | $\gamma_\mathrm{r}/2\pi$(MHz) | $\gamma_\mathrm{nr}/2\pi$(MHz) |
| 66.62 | 9.754 | 1.37 | 0.12 |

| Spin ensemble parameters | | | Coupling strengths |
|---|---|---|---|
| $\mu$ | $g_{\mathrm{L},\mu}$ | $\gamma_{\mathrm{s}\mu}/2\pi$(MHz) | $g_\mu/2\pi$(MHz) |
| 1 | 2.166 | 13.17 | 15.92 |
| 2 | 2.135 | 10.25 | 15.51 |
| 3 | 2.101 | 9.28 | 15.14 |
| 4 | 2.066 | 10.67 | 14.70 |
| 5 | 2.029 | 8.35 | 14.06 |
| 6 | 1.991 | 7.74 | 13.54 |
| 7 | 1.952 | 8.23 | 12.99 |
| 8 | 1.913 | 10.39 | 12.43 |

**Table S6:** $S_{11}(\omega)$ fitting parameters for VOTPP in position #B, corresponding to the dataset shown in Fig. S7 (d,e,f).

| Resonator parameters | | | |
|---|---|---|---|
| $a$(MHz/T$^2$) | $\bar{\nu}_0$(GHz) | $\gamma_r/2\pi$(MHz) | $\gamma_{nr}/2\pi$(MHz) |
| 70.93 | 9.795 | 1.46 | 0.15 |

| Spin ensemble parameters | | | Coupling strengths |
|---|---|---|---|
| $\mu$ | $g_{L,\mu}$ | $\gamma_{s\mu}/2\pi$(MHz) | $g_\mu/2\pi$(MHz) |
| 1 | 2.166 | 17.40 | 12.71 |
| 2 | 2.135 | 13.16 | 12.33 |
| 3 | 2.102 | 12.49 | 12.02 |
| 4 | 2.067 | 13.12 | 11.59 |
| 5 | 2.030 | 12.57 | 11.12 |
| 6 | 1.993 | 10.29 | 10.64 |
| 7 | 1.954 | 10.60 | 10.21 |
| 8 | 1.915 | 12.39 | 9.81 |

**Table S7:** $S_{11}(\omega)$ fitting parameters for VOTPP in position #C, corresponding to the dataset shown in Fig. S7 (g,h,i).

| Resonator parameters | | | |
|---|---|---|---|
| $a$(MHz/T$^2$) | $\bar{\nu}_0$(GHz) | $\gamma_r/2\pi$(MHz) | $\gamma_{nr}/2\pi$(MHz) |
| 65.81 | 9.818 | 1.59 | 0.56 |

| Spin ensemble parameters | | | Coupling strengths |
|---|---|---|---|
| $\mu$ | $g_{L,\mu}$ | $\gamma_{s\mu}/2\pi$(MHz) | $g_\mu/2\pi$(MHz) |
| 1 | 2.165 | 40.91 | 3.91 |
| 2 | 2.135 | 37.97 | 3.88 |
| 3 | 2.102 | 43.82 | 4.17 |
| 4 | 2.068 | 35.45 | 3.67 |
| 5 | 2.031 | 26.26 | 3.08 |
| 6 | 1.994 | 24.01 | 2.83 |
| 7 | 1.955 | 23.66 | 2.82 |
| 8 | 1.916 | 23.80 | 2.72 |

**Table S8:** $S_{11}(\omega)$ fitting parameters for VOTPP in position #D, corresponding to the dataset shown in Fig. 4 of main paper and in Fig. S7 (l,m,n).

| Resonator parameters | | | |
|---|---|---|---|
| $a$(MHz/T$^2$) | $\bar{\nu}_0$(GHz) | $\gamma_{\rm r}/2\pi$(MHz) | $\gamma_{\rm nr}/2\pi$(MHz) |
| 47.76 | 9.894 | 1.58 | 0.45 |

| Spin ensemble parameters | | | Coupling strengths |
|---|---|---|---|
| $\mu$ | $g_{{\rm L},\mu}$ | $\gamma_{{\rm s}\mu}/2\pi$(MHz) | $g_\mu/2\pi$(MHz) |
| 1 | 2.166 | 49.18 | 3.72 |
| 2 | 2.134 | 45.88 | 3.52 |
| 3 | 2.102 | 56.31 | 3.98 |
| 4 | 2.067 | 39.12 | 3.33 |
| 5 | 2.031 | 28.14 | 2.50 |
| 6 | 1.993 | 23.51 | 2.18 |
| 7 | 1.955 | 21.23 | 2.02 |
| 8 | 1.916 | 21.35 | 1.94 |



\end{document}